\documentclass[prd,aps,twocolumn,nofootinbib,preprintnumbers,amsfonts,showpacs,superscriptaddress]{revtex4-1}

\usepackage{amsmath} 
\usepackage{graphicx} 
\usepackage{amsthm}
\usepackage{amssymb} 
\usepackage{dsfont}
\usepackage{yfonts}
\usepackage{hyperref}
\usepackage{floatrow,dcolumn,rotating}
\usepackage{array,xcolor,graphicx}
\usepackage{booktabs,multirow}
\usepackage[utf8]{inputenc}
\usepackage{mathtools}

\usepackage[normalem]{ulem}

\hypersetup{colorlinks=true,linkcolor=blue,citecolor=red,urlcolor=blue}

\newcommand{\be}{\begin{equation}}
\newcommand{\ee}{\end{equation}}
\newcommand{\bea}{\begin{eqnarray}}
\newcommand{\eea}{\end{eqnarray}}

\newcommand{\nn}{\nonumber\\}

\def\la{\langle}
\def\ra{\rangle}

\def\CB{\mathcal{B}}
\def\CC{\mathcal{C}}

\def\CG{\mathcal{G}}
\def\CH{\mathcal{H}}

\def\CJ{\mathcal{J}}

\def\CM{\mathcal{M}}

\def\CO{\mathcal{O}}

\def\CV{\mathcal{V}}
\def\CW{\mathcal{W}}

\begin{document}
\preprint{MIT-CTP/4966}

\title{Generalised global symmetries in states with dynamical defects:\\ the case of the transverse sound in field theory and holography}
\author{Sa\v{s}o Grozdanov}
\email{saso@mit.edu}
\affiliation{Center for Theoretical Physics, MIT, Cambridge, MA 02139, USA}
\author{Napat Poovuttikul}
\email{nickpoovuttikul@hi.is}
\affiliation{Instituut-Lorentz for Theoretical Physics, Leiden University, Niels Bohrweg 2, Leiden 2333 CA, The Netherlands}
\affiliation{University of Iceland, Science Institute, Dunhaga 3, IS-107, Reykjavik, Iceland}
\begin{abstract}
In this work, we show how states with conserved numbers of dynamical defects (strings, domain walls, etc.) can be understood as possessing generalised global symmetries even when the microscopic origins of these symmetries are unknown. Using this philosophy, we build an effective theory of a $2+1$-dimensional fluid state with two perpendicular sets of immersed elastic line defects. When the number of defects is independently conserved in each set, then the state possesses two one-form symmetries. Normally, such viscoelastic states are described as fluids coupled to Goldstone bosons associated with spontaneous breaking of translational symmetry caused by the underlying microscopic structure---the principle feature of which is a transverse sound mode. At the linear, non-dissipative level, we verify that our theory, based entirely on symmetry principles, is equivalent to a viscoelastic theory. We then build a simple holographic dual of such a state containing dynamical gravity and two two-form gauge fields, and use it to study its hydrodynamic and higher-energy spectral properties characterised by non-hydrodynamic, gapped modes. Based on the holographic analysis of transverse two-point functions, we study consistency between low-energy predictions of the bulk theory and the effective boundary theory. Various new features of the holographic dictionary are explained in theories with higher-form symmetries, such as the mixed-boundary-condition modification of the quasinormal mode prescription that depends on the running coupling of the boundary double-trace deformations. Furthermore, we examine details of low- and high-energy parts of the spectrum that depend on temperature, line defect densities and the renormalisation group scale. 
\end{abstract}

\maketitle
\begingroup
\hypersetup{linkcolor=black}
\tableofcontents
\endgroup

\section{Introduction: states with generalised global symmetries}\label{sec:Introduction}

The enumeration of symmetries in a physical state of interest is a crucial step in the construction of an effective field theory that governs its dynamics. In particular, it is the global symmetries, not the local gauge symmetries of the underlying microscopic interactions, that are relevant for the behaviour of long-range excitations of that theory. A particular example that we will refer to throughout this work is magnetohydrodynamics (MHD) in $d + 1 = 4$ spacetime dimensions. MHD is an effective hydrodynamic theory of plasmas---gas-like states of matter with screened electromagnetic interactions. A purely symmetry based construction of MHD was recently presented in Ref. \cite{Grozdanov:2016tdf}, of which the philosophy we continue in this work. The essential ingredient is the concept of a generalised global symmetry \cite{Gaiotto:2014kfa}\footnote{See also Refs. \cite{Nussinov:2006iva,Nussinov:2017xxx,Kapustin:2013uxa,Kapustin:2014gua} for earlier studies of generalised (higher-form) global symmetries in the context of topological phases of matter.}, which we will review in the context of MHD and expand upon below.\footnote{Another symmetry-related ingredient in the choice of a state is the symmetry-breaking pattern of spacetime symmetries. In a relativistic fluid at finite temperature or density in $d+1$ spacetime dimensions, with $d \geq 3$, the preferred (non-relativistic) choice of a rest-frame breaks the Lorentz group as $SO(d,1) \to SO(d)$. The presence of an additional vector field, e.g. the magnetic field in MHD, further breaks $SO(d)\to SO(d-1)$. In $d=2$, such a symmetry breaking pattern only leaves a discrete $\mathbb{Z}_2$ unbroken. In MHD in $d=3$ at $T=0$, the symmetry is enhanced by a boost symmetry along magnetic field lines to $SO(1,1) \times SO(2)$ \cite{Grozdanov:2016tdf}.}

Relativistic MHD in $d=3$ without any externally applied sources possess a conserved stress-energy tensor $\nabla_\mu T^{\mu\nu} = 0$, which corresponds to conserved energy and momentum. Furthermore, a conserved number of magnetic flux lines crossing a (spacelike) codimension-two surface gives rise to a $U(1)$ generalised global symmetry with an associated conserved (antisymmetric) two-form current $\nabla_\mu J^{\mu\nu} =0$. Together, these two global symmetries can be used in the construction of an effective theory of MHD \cite{Grozdanov:2016tdf}.\footnote{See Ref. \cite{Hernandez:2017mch} for a construction of relativistic MHD in the language of electromagnetic fields, and relation to \cite{Grozdanov:2016tdf}. Related past works include \cite{Kovtun:2016lfw,Huang:2011dc,Critelli:2014kra,Finazzo:2016mhm,Montenegro:2017rbu}.} Physically, the conservation is a result of the absence of magnetic monopoles and in the standard language of Maxwell's electrodynamics, $J^{\mu\nu} = \frac{1}{2} \epsilon^{\mu\nu\rho\sigma} F_{\rho\sigma}$, where $F = dA$, with $A$ a one-form gauge field. The conservation of $J^{\mu\nu}$ is thus the (topological) Bianchi identity, $dF = 0$. In a state of matter with dynamical electromagnetism that contains massless photons---i.e. the electromagnetic spectrum is at least well approximated by the spectrum of operators near the vacuum state---the statement that $\nabla_\mu J^{\mu\nu} = 0 $ ($d^2 A = 0$) is a tautology. However, once this identity is understood as arising from a global symmetry, then one can make use of it in an effective theory of a complicated state with e.g. a hydrodynamic description, regardless of whether $A$ exists in the spectrum or not. In precisely this way, MHD can describe a plasma with gapped, massive photons, after we write $J^{\mu\nu}$ in a gradient expansion and treat $\nabla_\mu J^{\mu\nu} = 0 $ as a hydrodynamic conservation equation; a plasma state has no photons in the extreme infra-red (IR) part of the spectrum, which makes $A$ an inconvenient variable for writing down the effective theory \cite{Grozdanov:2016tdf}. A massless photon should then be thought of as a Goldstone boson of a spontaneously broken generalised global symmetry in a phase with condensed, tensionless magnetic flux lines (or strings). In this language, the Maxwell action $d A \wedge \star \, d A$ is an effective low-energy action in a (generalised global) symmetry broken phase, which non-linearly realises the symmetry $A \to A + d \alpha$. This is a higher-form analogue of the usual scalar Goldstone boson action $d\theta \wedge \star \, d\theta$ with the non-linearly realised shift symmetry, $\theta \to \theta + c$. Furthermore, we note that the order parameter which is able to distinguish between a broken and an unbroken symmetry is an expectation value of the 't Hooft loop that encircles the magnetic flux lines (or strings). In a symmetry preserving state, the expectation value obeys the area law---it scales as $\la W_C \ra \sim \exp \left\{-T \, \text{Area}[C] \right\}$. In the symmetry broken phase with a string condensate and massless photons, it obeys the perimeter law, $\la W_C \ra \sim \exp \left\{-T \, \text{Perimeter}[C] \right\}$ (see Ref. \cite{Gaiotto:2014kfa}). 

Beyond the microscopic, particle-oriented motivation for considering generalised global symmetries in effective theories, from the purely geometrical point of view, magnetic flux lines are one-dimensional strings and for this reason, MHD is equivalent to a theory of a string fluid, i.e. a hydrodynamic theory with a fluctuating conserved number of strings \cite{Grozdanov:2016tdf,Olesen:1995ff,Schubring:2014iwa}. The two theories lie within the same universality class of effective IR theories in the sense that they exhibit the same dynamics. This idea can be easily generalised to states with a conserved number of fluctuating $p$-dimensional topological defects (strings, branes), which have an associated conserved charge $Q = \int_S \star J$, with $S$ a $d-p$ dimensional surface and $J$ a $(p+1)$-form current. The symmetry is then called a $p$-form symmetry. Furthermore, one can imagine systems with multiple such conserved quantities and a $U(1) \times U(1) \times \ldots$ generalised global symmetry group and multiple higher-form conserved currents. Unlike in MHD, where the microscopic origin of the generalised global symmetry is clear, here, it is our goal to show the usefulness of such symmetries in constructions and geometrical understanding of field theories of states which readily appear in nature. 

\begin{figure}[h]
\begin{center}
\includegraphics[width=0.5\textwidth]{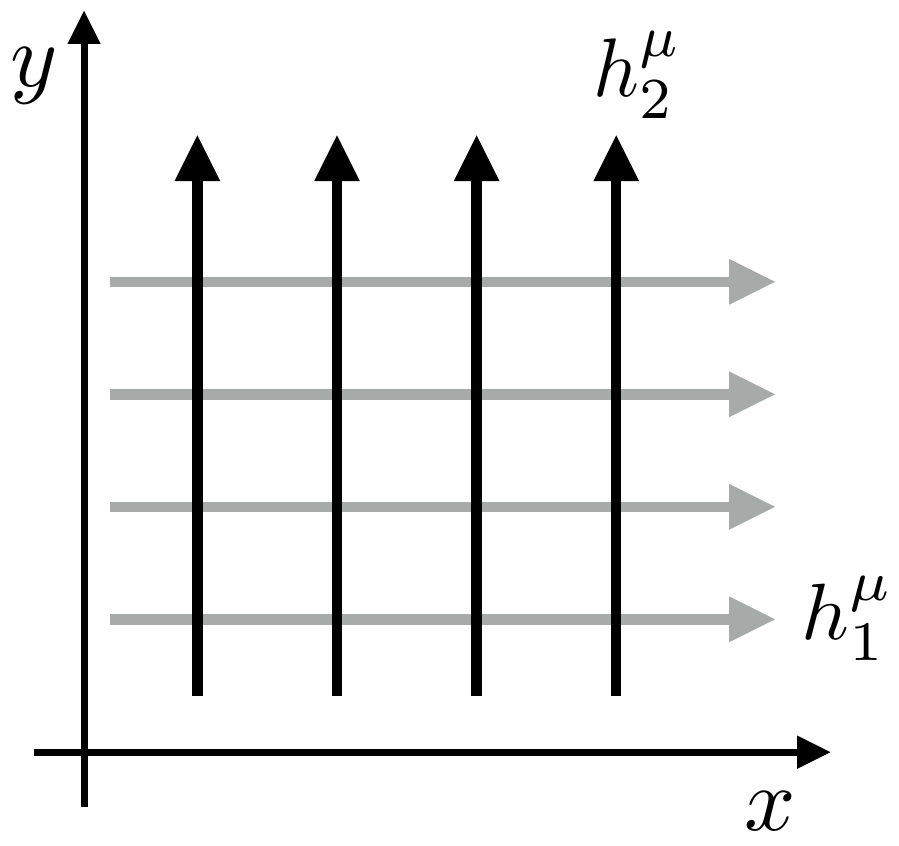}
\end{center}
\caption{The equilibrium configuration with two sets of perpendicular lines. The number of these fluctuating extended objects (defects) is conserved in each of the two mutually perpendicular sets of strings, which gives rise to two independent generalised global symmetries.}
\label{fig:strings}
\end{figure}

In this paper, we consider one of the simplest such cases: an isotropic $d=2$ state with two sets of independently conserved numbers of one-dimensional defects---i.e. two one-form symmetries. Beyond a conserved stress-energy tensor $T^{\mu\nu}$, the relevant generalised global symmetries in this state form a $U(1)\times U(1)$ group with each $U(1)$ characterised by a conserved two-form current $J_I$, for $I = \{1,2\}$:
\begin{align}
\nabla_\mu T^{\mu\nu} = 0\,, && \nabla_\mu J^{\mu\nu}_{I} = 0 \,.
\label{eq:wardTJ}
\end{align}
At the ideal hydrodynamic level (see Ref. \cite{Grozdanov:2016tdf}), we can write the constitutive relations for the three composite conserved operators as
\begin{align}
T^{\mu\nu} &= \left(\varepsilon + p \right) u^\mu u^\nu + p \,g^{\mu\nu} - \sum_{I=1}^2 \mu_I \rho_I h_I^\mu h_I^\nu \,,\label{eq:DefIdealStessEn} \\
J_I^{\mu\nu} & = 2 \rho_I u^{[\mu} h_I^{\nu]} \, ,\label{eq:DefIdealCurrent}
\end{align}
where the two spacelike vectors $h^\mu_I$ (both normalised to $h_{I,\mu} h^\mu_I = 1$) are macroscopic, hydrodynamic variables parametrising the equilibrium directions of extended objects charged under the two $U(1)$ one-form symmetries. Without loss of generality, they are chosen to be orthogonal to the fluid velocity, i.e. $u_\mu h^\mu_I = 0$. The absence of anti-symmetric $h_I^{[\mu} h_J^{\nu]}$ terms in \eqref{eq:DefIdealCurrent} is required by factorisation of the two generalised global symmetries. The scalar functions $\varepsilon$, $p$, $\mu_I$ and $\rho_I$ are the energy density, pressure, two chemical potentials and two densities of the line-defect fluxes, which satisfy the thermodynamic relations 
\begin{align}
\varepsilon + p &= s T + \sum_{I=1}^2 \mu_I \rho_I \,, \label{Thermo1} \\
dp &= s \, dT + \sum_{I=1}^2 \rho_I d\mu_I \, . \label{Thermo2}
\end{align}
We consider the scenario whereby in equilibrium, the two sets of strings are perpendicular to each other, $h_{1,\mu}h^\mu_{2} = 0$, and for simplicity, we choose the vectors $h^\mu_{I}$ to point in $x$ and $y$ directions for $I=\{1,2\}$, respectively (see Fig. \ref{fig:strings}). The construction of the above theory of ideal hydrodynamics from the point of view of the equilibrium partition function is presented in Appendix \ref{app:EQPF}. For the purposes of this work, we will not need the classification of the constitutive relations \eqref{eq:DefIdealStessEn} and \eqref{eq:DefIdealCurrent} at higher orders in the gradient expansion.\footnote{For details on how one systematically classifies higher-order terms in a theory of hydrodynamics, see e.g. \cite{Baier:2007ix,Romatschke:2009kr,Kovtun:2012rj,Grozdanov:2015kqa}.}

What we will argue here is that as a string fluid is equivalent to MHD, which is itself an effective theory of the dynamics of plasmas with some complicated microscopic quantum description, the setup studied here is equivalent to a theory of long-range excitation of fluids with additional properties of an elastic medium---a viscoelastic medium.\footnote{More precise definitions of viscoelasticity vary in the literature. In this work, we use the word viscoelasticity to refer to any combination of fluid- and solid-like properties.} While the effective theory discussed so far has been insensitive to detailed physics at the scales of any potentially underlying microscopic crystalline or lattice structure immersed inside the fluid, the symmetries, which incorporate the existence of line defects, and the assumption that the infra-red (IR) limit is sufficiently well described by a (hydrodynamic) gradient expansion of the relevant fields can reproduce known features of elastic media, such as transverse sound modes (transverse phonon excitations). The analysis of these transverse excitations, which are normally associated with Goldstone modes arising from spontaneously broken spacetime symmetries (broken translations by a lattice), will be our main focus in this work.\footnote{We thank Jan Zaanen for numerous discussions regarding the nature of solids and elasticity theory.} 

In analogy with the absence of spontaneous symmetry breaking of standard continuous global (zero-form) symmetries in $d=1$, a generalised version of the Coleman-Mermin-Wagner theorem states that a continuous $p$-form symmetry cannot be broken when $d - p < 2$ \cite{Gaiotto:2014kfa}.\footnote{We will not discuss the possibility of a potential BKT-type phase transition in the breaking of higher-form symmetries in this work.} For this reason, one-form symmetries in $d=2$, which are of interest to this work, cannot be spontaneously broken. However, a two-form current $J$ in $d=2$ can still be expressed in terms of a topological identity of a ``magnetic" symmetry, as in the case of electrodynamics in $d=3$. In $d=3$, $J$ satisfies the Bianchi identity $d \star  J = 0$ in both broken and unbroken phases. In terms of a photon field, it takes the form $J = \star \, d A$. Similarly, in $d=2$, two-form currents of the magnetic one-form symmetries that satisfy a Bianchi-type identity $d \star  J_I = 0$ can be written as
\begin{align}\label{ScalarHodge1}
d \star J_I = d \xi_I = 0 \, ,
\end{align}
where $\xi_I$ are the one-form analogues of the electromagnetic field strength that, expressed as closed and exact forms, become
\begin{align}\label{ScalarHodge2}
\xi_I = d \psi_I \,,
\end{align}
where $\psi_I$ are some zero-form (scalar) fields. Imagine now that in the absence of matter---in our case an absent fluid component of the state---there also exists an ``electric" symmetry, $d\star \xi_I = 0$, which is a zero-form symmetry with a one-form current $\xi_I$. Expressions \eqref{ScalarHodge1} and \eqref{ScalarHodge2} then imply that the scalars $\psi_I$ are massless and obey the equation $\partial_\mu \partial^\mu \psi_I = 0$, which enables us to think of $\psi_I$ as the Goldstone bosons of a spontaneously broken ``electric" symmetry. However, in the presence of matter (the fluid), depending on the microscopic details of the theory, this symmetry could become explicitly broken and all fields in the theory generically gapped.\footnote{In a superfluid, a global zero-form $U(1)$ symmetry is spontaneously broken and the associated one-form current is a function of both the matter sector and the zero-form Goldstone boson.}

From a point of view that is independent of the above discussion of zero- and one-form symmetries, the introduction of another set of massless scalar fields, which we call $\phi_I$, can be used to construct an effective action of a fluid with spontaneously broken translational invariance or a pure solid without the fluid component \cite{kleinert1989gauge,Beekman:2016szb,Endlich:2012pz,Nicolis:2013lma,Delacretaz:2014jka}. The physical origins of the $\phi_I$ and $\psi_I$ fields are different. More precisely, $\phi_I$ arise as a result of a spontaneous breaking of translations and may only be invoked in reference to an explicit presence of a lattice or a crystalline microscopic structure in the state. On the other hand, the existence of $\psi_I$ is in no way predicated upon broken translational symmetry and these fields may exist in a theory with translational invariance. As will be shown in Section \ref{sec:HydroVsElasticity}, however, the two effective actions (partition functions) with $\psi_I$ and $\phi_I$ can be equal. Hence, two independent points of view can lead to equivalent effective actions. As already noted above, when the fluid component is present, the situation is more complicated and will depend on the microscopic details of the state. We will further elaborate on these issues in Section \ref{sec:HydroVsElasticity}. For now, we only note that as we will argue in Section \ref{sec:HydroVsElasticity}, a more precise connection between the standard theory of elasticity (with $\phi_I$) and a one-form symmetry-based description can be made by realising that the elastic theory itself possesses two (magnetic) one-form symmetries, with conserved two-form currents that are equivalent to $J_I$. Furthermore, and most importantly, we will show that both descriptions of a viscoelastic material are equivalent, at least at the linear level studied in this work. 

To conclude the introductory discussion, we note that as the Bianchi identity in $d=3$ electromagnetism embodies the absence of monopoles, the conservation of $J^{\mu\nu}_I$ in \eqref{eq:wardTJ} in the language of elasticity theory corresponds to the absence of dislocations in a lattice of lines in Fig. \ref{fig:strings} (see Section \ref{sec:HydroVsElasticity}). Furthermore, as in MHD \cite{Grozdanov:2016tdf}, an external two-form field modifies the conservation of the stress-energy tensor to 
\begin{align}\label{ModWardJ}
\nabla_\mu T^{\mu\nu} = \sum_{I=1}^2 H^{\nu}_{I,\rho\sigma} J^{\rho\sigma}_I \,,
\end{align}
where $H_I = d b_I$ are three-form field strengths of the two-form fields $b_{I,\mu\nu}$ that can be used to gauge or source the conserved two-form operators. Treating the identities \eqref{eq:wardTJ} (and \eqref{ModWardJ}) as genuine global conservation laws allows us to not only construct a hydrodynamic, low-energy effective theory, but by following \cite{Grozdanov:2017kyl,Hofman:2017vwr} to also build a holographic bulk dual in which these symmetries and their associated Ward identities are manifest. As we will show in Section \ref{Sec:Holography}, the dual embeds a low-energy viscoelastic state into a theory with additional high-energy excitations. The simplest incarnation of the statement that the holographic dual does not describe a pure solid but a viscoelastic state follows from considering the state at zero density of line defects, $\rho_I = 0$. In that case, as we will see in Section \ref{sec:zerowall}, the system will exhibit diffusive properties of a standard dissipative fluid. In fact, as is usual in holography, the fluid component of the state remains particularly robust.\footnote{Holographic setups without pathologies are unconducive to descriptions of solids or insulators even in the presence of strong disorder \cite{Grozdanov:2015qia,Grozdanov:2015djs} due to universal properties associated with event horizons (see also \cite{Donos:2015gia,Banks:2015wha,Donos:2017ihe}).} At finite density, $\rho_I >0$, the system will exhibit solid-like features, such as the transverse sound mode. We note that in contrast to holographic setups in which the theory is engineered to have a spatially modulated phase transition at low temperature \cite{Ooguri:2010kt,Donos:2012wi,Donos:2011bh,Rozali:2012es,Donos:2013gda,Withers:2013loa,Jokela:2016xuy,Cremonini:2016rbd,Cai:2017qdz,Andrade:2017cnc}, properties of an elastic material, which will arise in perturbations around a homogeneous and isotropic background, are in this work a consequence of global symmetries alone.

It should also be made clear that this work is not intended as an exploration of new properties of fluids or solids, but as a demonstration of the versatility of the language of effective field theories with generalised global symmetries, both from the point of view of field theory and holography. The phenomenology of the hydrodynamic constitutive relations and conservation laws presented in this section---i.e. the effective field theory approach---will be examined in Section \ref{sec:EFT}. In Section \ref{Sec:Holography}, we then extend the discussion of the holographic duals of states with generalised global symmetries which were first constructed in Refs. \cite{Grozdanov:2017kyl,Hofman:2017vwr} (and applied to MHD), and study the simplest, analytically tractable holographic dual of the system of interest here. 

\paragraph*{Note added:} While this manuscript was being completed, we became aware of the works in Refs. \cite{Alberte:2017oqx,Amoretti:2017frz}, which contain some overlap with our analysis of transverse sound modes.

\section{Effective field theory}\label{sec:EFT}

We begin our study of the theory set up in Section \ref{sec:Introduction} by analysing its collective excitations. Of main focus will be the transverse sector of the retarded two-point correlation function spectrum of conserved operators---i.e. the linearised, long-lived excitations. The reason for focusing on the transverse sector is that its behaviour can be used to distinguish between fluid- and solid-like properties of states. In the former case, one finds diffusive modes and in the latter case, propagating sound waves. The longitudinal channel on the other hand always exhibits sound modes. Beyond the exploration of the properties of the linear spectrum, in the second part of this section, we will make a closer comparison between the theory from Section \ref{sec:Introduction} and a theory of hydrodynamics with spontaneously broken translational symmetry, as well as elasticity theory.

\subsection{Transverse fluctuations}\label{sec:LineFieldTheory}

The spectrum of the retarded two-point stress-energy tensor correlators in the transverse channel depends on the coupled set of linearly perturbed hydrodynamic fields, $T$, $\mu_I$, $u^\mu$ and $h^\mu_I$, which, for a momentum chosen to point in the $x$-direction, are odd under the parity transformation $y \to - y$.  Analogous results for the longitudinal channel are presented in Appendix \ref{appendix:Longitudinal}. The equilibrium state of interest to this work is characterised by an isotropic and homogeneous choice of the line defect densities and chemical potentials,
\begin{align}
\mu_{1} = \mu_{2} =  \mu\,,&&  \rho_{1} = \rho_{2} = \rho \,.
\end{align}
We note that these conditions could easily be relaxed if one was interested in exploring the dynamics of more complicated states.

At the ideal (non-dissipative) level of the constitutive relations \eqref{eq:DefIdealStessEn} and \eqref{eq:DefIdealCurrent}, we find that the equations of motion \eqref{eq:wardTJ} reduce to the following system:\footnote{The $t$-components of the two conservation equations $\nabla_\mu J^{\mu\nu}_{I} = 0$ give algebraic relations between $\delta h^\mu_{I}$ and $\delta \mu_{I}$. For transverse perturbations, we find that $\delta h_{2}^x=0$ and that $\delta \mu_{2}$ can be expressed in terms of $\delta T$ and $\delta \mu_{1}$.}
\begin{align}
\partial_t \delta u_y - \CV_A^2 \partial_x \delta h_1^y &=  0 \,,\label{eq:eom2Form1}\\
\partial_t \delta h^y_1 -\partial_x \delta u_y &= 0 \,, \label{eq:eom2Form2}
\end{align}
where 
\begin{align}
\CV_A^2 = \frac{ \mu\rho}{ \varepsilon+p-\mu\rho } \, .
\end{align} 
A Fourier decomposed plane-wave solution of Eqs. \eqref{eq:eom2Form1} and \eqref{eq:eom2Form2} proportional to $e^{-i \omega t + i k x}$ is a propagating linear sound mode with the dispersion relation
\begin{align}\label{DispRelSoundTrans}
\omega = \pm \CV_A k  + \CO(k^2) \, , 
\end{align}
which is analogous to the Alfv\'{e}n wave in MHD (see Ref. \cite{Grozdanov:2016tdf}).

By using the procedure of Kadanoff and Martin \cite{Kadanoff} (see also \cite{Kovtun:2012rj}), one can then find the leading-order hydrodynamic approximation to the full retarded two-point function of a transverse stress-energy tensor,
\begin{equation}\label{eq:TtyTtyCorrelator}
G^{ty,ty}_{TT,R}(\omega,k) =  \frac{\mu\rho \, k^2}{\omega^2-\CV_A^2k^2} \,,
\end{equation}
which can be used to find both the sound pole with the dispersion relation \eqref{DispRelSoundTrans} and the residue. The other transverse correlator with the same pole structure, $G^{xy,xy}_{TT,R}$, can be found by using the Ward identity $k_\mu G^{\mu\nu,\rho\sigma}_{TT,R} = 0$. $G^{xy,xy}_{TT,R} $ has a non-zero real part at vanishing momentum, which will be of importance below, and follows from the following expression:
\begin{equation}\label{eq:TxyTxyCorrelator}
G^{xy,xy}_{TT,R}(\omega,k=0) = \mu\rho +\CO(\omega) \,.
\end{equation}
It is important to keep in mind that the two-point functions computed using the procedure of \cite{Kadanoff} do not include the contact terms, i.e. all contact terms are subtracted. Furthermore, we note that due to the coupling of the hydrodynamic degrees of freedom at non-zero density, the sound poles of \eqref{eq:TxyTxyCorrelator} are also the poles of the transverse $\la J^{\mu\nu}_1J_1^{\rho\sigma}\ra_R$ two-point functions. In total, the transverse hydrodynamic sector of the full theory with our specific choice of the momentum pointing in the $x$-direction contains two modes---a pair of sound modes. The choice of the polarisation of the momentum decouples the fluctuations of the $h^\mu_2$ field, which can be set to zero.   

For comparison, we note that the above features of the transverse channel are markedly different from those in uncharged and charged (under a zero-form symmetry) relativistic fluids without a generalised global symmetry and without a conserved two-form current $J^{\mu\nu}$. There, the (first-order dissipative) hydrodynamic pole of both $G^{ty,ty}_{TT,R}$ and $G^{xy,xy}_{TT,R}$ is a diffusive mode $\omega = -i D k^2$ and to leading order at small $\omega$, the equilibrium contact-term subtracted $G^{xy,xy}_{TT,R} (\omega, k = 0) = - i\omega \eta$, where $\eta$ is the shear viscosity.\footnote{In general, correlation functions depend on contact terms. The hydrodynamic correlator $G^{xy,xy}_{TT,R} (\omega, k)$, including contact terms, can be obtained by varying $T^{xy}$ with respect to the linear background metric perturbation $\delta h_{xy}$. This procedure gives the equilibrium pressure in the following limit: $\lim_{\omega \to 0 } G^{xy,xy}_{TT,R} (\omega, k = 0) = p$ (see Ref. \cite{Kovtun:2012rj}).} The dispersion relation and the $G^{xy,xy}_{TT,R} $ have no ideal hydrodynamic contributions. From the point of view of the theory of elasticity, the sound pole of $G^{ty,ty}_{TT,R}(\omega,k)$ in \eqref{eq:TtyTtyCorrelator} and the non-vanishing real part of $G^{xy,xy}_{TT,R}(\omega,k=0)$ in \eqref{eq:TxyTxyCorrelator} can be interpreted as the dispersion relation of a transverse phonon and as the shear elastic modulus, respectively (see e.g. \cite{Alberte:2015isw,Amoretti:2017frz}). 

Lastly, we note that the longitudinal channel contains propagating sound modes (see Appendix \ref{appendix:Longitudinal}), as do gases, fluids and plasmas. This property makes the longitudinal channel a less useful diagnostic tool for distinguishing solids or viscoelastic states from other fluid-like states of matter.  

\subsection{Hydrodynamics with spontaneously broken translational symmetry and the theory of elasticity}\label{sec:HydroVsElasticity}

Phonons are Goldstone bosons associated with spontaneous symmetry breaking of translational symmetry \cite{Leutwyler:1996er}. Usually, they are employed to describe a theory of long-range excitations in solids with a dynamically formed microscopic lattice structure---details of which one expects the effective IR theory to be insensitive to at length scales much longer than the lattice spacing. In this extreme IR limit, only the symmetries and the identification of long-range modes distinguishes a fluid from a solid (see e.g. \cite{Eshelby1956,Son:2000ht,Dubovsky:2011sj,Endlich:2012pz,Nicolis:2013lma}). Indeed, this is the sense in which we interpret our theory from Section  \ref{sec:Introduction} to contain the knowledge of well-known elastic properties such as a transverse sound mode seen in Section \ref{sec:LineFieldTheory}. In this section, we show further supporting evidence for making this claim.

Let us for now focus only on the elastic theory of a pure solid. In its simplest form, the displacements of the crystalline structure can be parametrised in terms of Goldstones $\phi_I$ with an action (see Ref. \cite{kleinert1989gauge,Beekman:2016szb})
\begin{align}\label{ElasticityAction}
S = -\sum_{I,J = 1}^2 \int d^3 x \, C^{\mu\nu}_{IJ} \partial_\mu \phi_I \partial_\nu \phi_J \,,
\end{align}
where $C^{\mu\nu}_{IJ}$ is a state-specific tensor, which contains the information about the elastic moduli. The theory possesses two zero-form global symmetries, $d\star P_I = 0$, with the one-form currents $P_I$---i.e. the momenta of the solid---given by $P^\mu_I = -C^{\mu\nu}_{IJ} \partial_\nu \phi_J$. As already discussed in Section \ref{sec:Introduction}, a priori, these fields cannot be directly equated with the Goldstone bosons of spontaneously broken electric zero-form symmetries. However, both lines of reasoning can lead to equal effective actions. Thus, the elastic theory should be thought of as a specific example of theories that can be constructed from considerations of Section \ref{sec:Introduction}. In fact, a case in which the compact $U(1) \times U(1)$ electric zero-form symmetry currents could be directly related to $P_I$ is one in which an IR state exhibits a periodic potential---a lattice. More precisely, the $U(1)$ group is isomorphic to the quotient group $\mathbb{R} / \mathbb{Z}$, i.e. $U(1) \cong \mathbb{R} / \mathbb{Z}$. Such a structure imposes a periodic identification of the microscopic fields, e.g. of some field $\Phi$, so that $\Phi(\phi_I) = \Phi(\phi_I + \ell)$, where $\ell$ is the lattice spacing. By setting the elastic tensor $C^{\mu\nu}_{IJ}$ to take a special form required by our choice of the state, the Goldstones $\phi_I$ can then be identified with $\psi_I$.    

In the presence of matter, the situation becomes less transparent and strongly dependent on the microscopic details of the UV theory. Imagine coupling $\psi_I$ to a matter sector in a way that the zero-form symmetries are explicitly broken. In that case, the state can still break (spatial) translational symmetry and $\phi_I$ can remain massless, whereas the Goldstones $\psi_I$ would in the case of an explicitly broken electric $U(1)\times U(1)$ symmetry become massive. 

What is more important for the present discussion is the fact that the action \eqref{ElasticityAction} also gives rise to two global magnetic one-form symmetries of Eqs. \eqref{ScalarHodge1} and \eqref{ScalarHodge2}, with $\psi_I$ replaced by $\phi_I$:
\begin{align}\label{twoformelast}
J_I^{\mu\nu} = \epsilon^{\mu\nu\rho} \partial_\rho \phi_I \, .
\end{align}
The essential requirement, which ensures the existence of the symmetry, i.e. that $\partial_\mu J^{\mu\nu}_I = 0$, is that the fields $\phi_I$ are single-valued, or that $\partial_{[\mu} \partial_{\nu]} \phi_I = 0$. This condition precisely encodes the absence of dislocations in a conventional language of elasticity theory (see e.g. \cite{Beekman:2016szb}). Hence, at least in the absence of the fluid component of the state, the IR limits of the two theories have the same symmetries. It should be noted that as in the case of electrodynamics with a massless photon field $A$, the symmetry in equation \eqref{twoformelast} is a tautological identity due to the fact that $J^{\mu\nu}_I$ is expressed in terms of $\phi_I$. On the other hand, however, the symmetry-based formalism of Section \ref{sec:Introduction} does not require the existence of any such field (neither $\phi_I$ nor $\psi_I$) and can be used to describe a variety of different states. 

Moving beyond the case of a pure solid and our discussion of microscopic realisations of different symmetries, we proceed with a phenomenological discussion of viscoelastic fluids, which are states with spontaneously broken translational invariance that exhibit both fluid- and solid-like properties. As already discussed in Section \ref{sec:Introduction}, in the effective description, it is standard to combine a theory of the two massless scalar fields $\phi_I$ with a hydrodynamic theory that includes the following (neutral) degrees of freedom: $u^\mu$ and $T$ (see \cite{chaikin2000principles,PhysRevA.6.2401,PhysRevB.22.2514} and also \cite{Delacretaz:2017zxd}). In such a state, the expectation values for the two $\phi_I$ are taken to be $\la \phi_1 \ra = \alpha  x $ and $\la \phi_2 \ra = \alpha  y$. Then, one can expand $\phi_I$ as 
\begin{align}\label{GoldstonesEquil}
\phi_I  = \alpha \left(  \delta^i_{~I} \, x_i + \pi_I \right) \, ,
\end{align}
and instead of $\phi_I$ think of the two fields $\pi_I$ as the (linearised) Goldstone bosons of spontaneously broken spatial symmetries---i.e. the phonons that describe the elastic modes of a solid structure---or consider $\phi_I$ as the fields that realise the symmetry non-linearly (see Ref. \cite{Nicolis:2013lma}). Since the symmetry is broken spontaneously, the Ward identity (energy-momentum conservation) of a fluid with the hydrodynamic $u^\mu$ and $T$ fields is modified to
\begin{equation}\label{EoM1Mod}
\partial_\mu T^{\mu\nu}(u^\mu,T,\phi_I) = 0 \,,
\end{equation}
with the spatial part of stress-energy tensor (the stress tensor) written as
\begin{align}
T_{ij} =&\, p_0 \,\delta_{ij}  -(\CB+\CG) (\partial_k  \phi^k) \,\delta_{ij} \nn
& - 2\CG \left[ \partial_{(i}\phi_{j)} -  (\partial_k \phi^k ) \, \delta_{ij}\right] , \label{eq:stressElastic}
\end{align}
where $\CB$ and $\CG$ are the bulk and the shear moduli of the isotropic solid component of the system, respectively. The expression in Eq. \eqref{eq:stressElastic} is obtained by combining the pressure term and the spatial component of the momentum $T^{ij} = P_I^i \delta^{Ij}$, where the spatial component of the elastic tensor $C^{\mu\nu}_{IJ}$ for an isotropic system is positive-definite and can be decomposed into \cite{chaikin2000principles,kleinert1989gauge}
\begin{equation}
C^{ij}_{IJ} = \CB \,\delta^{ij}\delta_{IJ}  + \CG \left( \delta^i_I\delta^j_J + \delta^i_J\delta^j_I - \delta^{ij}\delta_{IJ} \right).
\end{equation}
The parameter $p_0$ is not the thermodynamic pressure $p$, only the fluid component of the total pressure.\footnote{For a derivation of the constitutive relation \eqref{eq:stressElastic} in the presence of external transtional symmetry breaking, see \cite{Bhattacharyya:2008ji,Blake:2015epa,Blake:2015hxa,Burikham:2016roo}. In analogy with the presently studied case with spontaneous symmetry breaking, the parameter $p_0$, as is conventionally chosen, does not take into account the contribution from kinetic terms of the scalar fields and is not equal to the thermodynamic pressure of the system.} Note also that we have projected the ``flavour index" $I$ to the spatial index $i$. In this notation, the elastic strain tensor is defined as $U_{ij} \equiv \partial_{(i}\phi_{j)}$ and $\delta U_{ij} = \partial_{(i} \pi_{j)}$ as the linearised fluctuation of the tensor around the equilibrium configuration, cf. Eq. \eqref{GoldstonesEquil}. We also note that as already discussed around Eq. \eqref{ModWardJ}, the conservation equation \eqref{EoM1Mod} encodes the absence of dislocations.

What is now the status of the magnetic one-form symmetries of the elasticity theory \eqref{ElasticityAction}? We claim that as in the case of electrodynamics in $d=3$, the global one-form symmetries remain unbroken in a viscoelastic state. Even though both descriptions include these symmetries, which should control their IR behaviour, the two one-form symmetries are normally not considered neither in the theory of elasticity nor viscoelasticity. To make contact between the two points of view, we first notice that the modified hydrodynamic theory (with $\phi_I$'s) has more field variables than equations of motion, which requires one to impose an equation connecting $\phi_i$ to the fluid variables. As the construction is equivalent to the two-fluid model of superfluidity, the equation that is chosen to connect them in the transverse channel is the analogue of the Josephson relation (see e.g. Chapter 8 of \cite{chaikin2000principles} or \cite{Delacretaz:2017zxd}):\footnote{In the two-fluid Landau-Tisza model of superfluids, the Josephson relation reads $u^\lambda \xi_\lambda = \mu$, where $\xi_\lambda \equiv \partial_\lambda \phi$ is the superfluid velocity, $\phi$ is the Goldstone boson of a spontaneously broken global $U(1)$ symmetry and $\mu$ the chemical potential (see e.g. \cite{Son:2000ht,Pujol:2002na,Bhattacharya:2011tra}).}
\begin{equation}\label{eq:JosephsonCon}
\partial_t \pi_i = \delta u_i \,,
\end{equation}
where $u_i$ are the spatial components of the fluid velocity. The system of equations is now closed, which allows us to solve it. In the language of one-form symmetries, the relation \eqref{eq:JosephsonCon} follows from matching the ``microscopic" description of a solid (in terms of $\phi_I$) with the effective description written in terms of the hydrodynamic fields. More precisely, we use the ideal part of the gradient expansion of $J^{\mu\nu}_I$ from \eqref{eq:DefIdealCurrent} and equate it to $J^I_{\mu\nu}$ computed from \eqref{ElasticityAction}. In the transverse channel, from the $xy$-components of $J^{\mu\nu}_I$, we find
\begin{align}
\delta J^{xy}_I = 2 \rho_I \delta u^{[x} h_I^{y]}   = \mathcal{K} \epsilon^{xy \lambda} \partial_\lambda \delta \phi_I \,,
\end{align}
which reproduces Eq. \eqref{eq:JosephsonCon} after an appropriate choice of the proportionality constant $\mathcal{K}$. Thus, in the language of generalised global symmetries of Section \ref{sec:Introduction}, the Josephson relation \eqref{eq:JosephsonCon} arises as a natural consequence of a global symmetry and does not need to be imposed independently. In terms of equation counting, there, $u^\mu$, $T$, $\mu_I$ and $h^\mu_I$ constitute in total $7$ degrees of freedom while Eq. \eqref{eq:wardTJ} na\"{i}vely gives $9$ differential equations. However, the two equations $\nabla_\mu J^{\mu t}_I = 0$ are constraints, which allows for the system to be closed. In the transverse channel, the two constraints directly play the role of the (derivative of the) Josephson relation \eqref{eq:JosephsonCon}. In the sound channel (see Appendix \ref{appendix:Longitudinal}), writing down the Josephson relation in its conventional form requires us to combine the constraints with a dynamical equation. Finally, it is important to note that because of this clear symmetry-related origin of the Josephson relation, dissipative corrections to the ideal relation also follow systematically from the hydrodynamic constitutive relations.\footnote{We also note that the Josephson relation in the transverse channel is in the sense of equation-counting equivalent of the magnetic Gauss's law in $3+1$-dimensional electrodynamics, $\vec \nabla \cdot \vec B  = 0$, which also follows from $\nabla_\mu J^{\mu t} = 0$.}  

\begin{figure}[h]
\begin{center}
\includegraphics[width=0.6\textwidth]{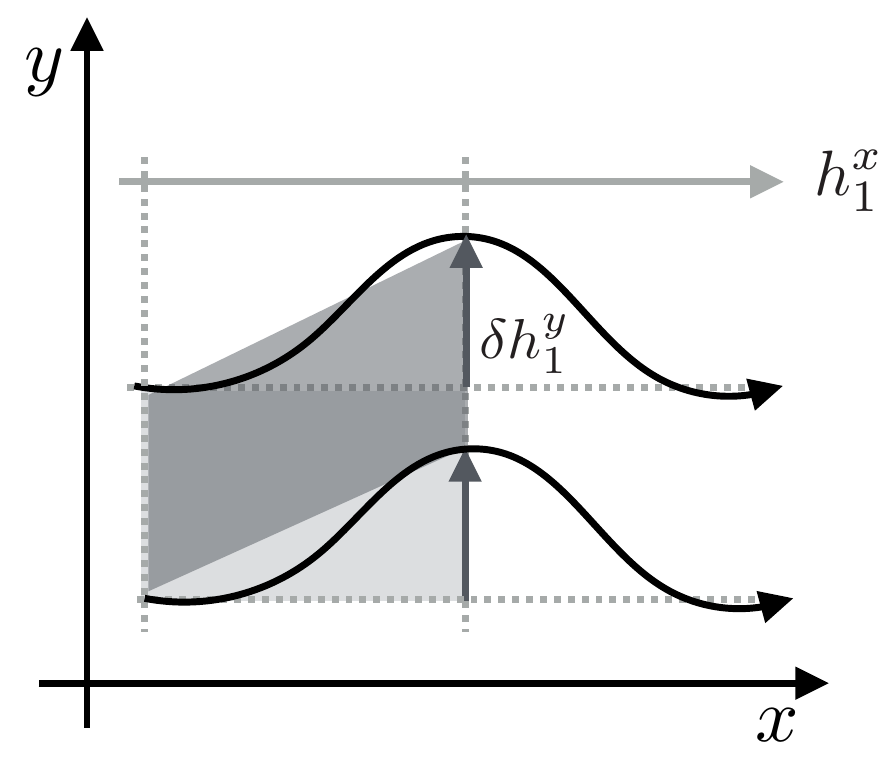}
\end{center}
\caption{A depiction of a transverse deformation of the line defect structures (a transverse lattice displacement) used to construct our theory in Section \ref{sec:Introduction}, which is parametrised by $\delta h^y_1$. In the language of conventional elasticity theory, this a shear deformation of the lattice, i.e. the $\delta U_{xy}$ perturbation.}
\label{fig:Deformstrings}
\end{figure}

Now, to show how the two descriptions match more precisely in the transverse channel, the relevant set of linearised differential equations that can be derived from Eqs. \eqref{EoM1Mod}, \eqref{eq:stressElastic} and \eqref{eq:JosephsonCon} is 
\begin{align}
 \chi_{p_yp_y}  \partial_t \delta u_y -2\CG \, \partial_x \delta U_{xy} & = 0 \,,\label{eq:eomElastic1} \\
2 \,\partial_t \delta U_{xy} - \partial_x \delta u_y & = 0 \,\label{eq:eomElastic2},
\end{align}
where $\chi_{p_yp_y}$ is the transverse momentum susceptibility. It follows immediately that the derivative of the Josephson relation \eqref{eq:eomElastic2} is identical to the equation derived from the conservation of the number of line defects \eqref{eq:eom2Form2} upon identifying $\delta h_1^y = 2 \delta U_{xy}$. One also notices that these two variables, $\delta h_1^y$ and $\delta U_{xy}$, describe the same physical deformation of the underlying ``lattice" structure. See Fig. \ref{fig:Deformstrings} for an illustration. Moreover, the identification of the one-point function $\la T^{xx}\ra$ and the real part of the retarded two-point function $G^{xy,xy}_{TT,R}$ in the two languages implies that 
\begin{align}\label{eq:Map2FormToElastic}
p - \mu\rho = p_0 - 2\CB \, ,&& \mu\rho = \CG \,.
\end{align}
Thus, the above mapping between degrees of freedom and variables used in the equations of state makes the systems of equations \eqref{eq:eom2Form1}--\eqref{eq:eom2Form2} and \eqref{eq:eomElastic1}--\eqref{eq:eomElastic2} identical. 

We can perform a similar matching between the formalism of Section \ref{sec:Introduction} and the theory of elasticity also in the longitudinal channel. Details are presented in Appendix \ref{appendix:Longitudinal}.

Finally, we note that while we have shown that a theory with generalised global symmetries can exhibit IR properties of an elastic medium, we leave a more detailed comparison with various precise incarnations of non-linear elasticity theory and its extensions to more complicated topological (lattice) structures to future works.

\section{Holography}\label{Sec:Holography}

We now turn our attention to constructing a holographic dual of a state with a conserved stress-energy tensor and two independently conserved two-form currents, cf. Eq. \eqref{eq:wardTJ}, and verifying the consistency of its IR properties with the effective field theory studied in Sections \ref{sec:Introduction} and \ref{sec:EFT}. This part is a direct extension of the works in Refs. \cite{Grozdanov:2017kyl,Hofman:2017vwr}. Here, however, we will work with a spatially isotropic black brane in one lower dimension, i.e. four spacetime bulk dimensions, which will make our discussion significantly simpler than in the case of a dual of MHD with a non-zero magnetic field \cite{Grozdanov:2017kyl} and partially analytically tractable. 

\subsection{The bulk theory and holographic renormalisation}\label{sec:HolRG}

We seek a four-dimensional bulk theory that is holographically dual to a state with a conserved stress-energy tensor $T^{\mu\nu}$ and two conserved two-form currents $J^{\mu\nu}_I$, $I \in \{1,2\}$. As in \cite{Grozdanov:2017kyl,Hofman:2017vwr}, it is simplest to consider a two-derivative theory with a fluctuating metric $G_{ab}$ and two decoupled two-form bulk fields $B_{I,ab}$, which source the three dual conserved operators via the following terms in the boundary generating functional: $\int d^3 x \, T^{\mu\nu} g_{\mu\nu}$ and $\sum_{I = 1}^2 \int d^3 x \, J^{\mu\nu}_I b_{I,\mu\nu}$. The holographic dictionary relating the  boundary sources $g_{\mu\nu}$ and $b_{I,\mu\nu}$ to the bulk $G_{ab}$ and $B_{I,ab}$ will be made precise below. 

The bulk action that we propose as a dual of a theory with two $U(1)$ one-form generalised global symmetries introduced in Section \ref{sec:Introduction} is 
\begin{equation}\label{eq:action1}
S = \frac{1}{2 \kappa_4^2} \int d^4 x \sqrt{-G} \left( R + \frac{6}{L^2} -\frac{1}{12} \sum_{I} H_{I,abc}H_{I}^{abc} \right) ,
\end{equation}
where $H_I=dB_I$. Henceforth, for convenience, we will set the Newton's constant in a way that sets $2 \kappa_4^2 = 1$ and fix the anti-de Sitter radius to $L=1$. 

The theory \eqref{eq:action1} has an isotropic, asymptotically anti-de Sitter black brane solution with the metric tensor
\begin{equation}\label{eq:DefMetric}
\begin{aligned}
ds^2 &= \frac{dr^2}{r^2f(r)} +r^2 \left(-f(r) dt^2 + dx^2+dy^2  \right),  \\
f(r) &= 1-\frac{m^2}{2r^2} - \left(1-\frac{m^2}{2r_h^2}\right)\frac{r_h^3}{r^3} \,,
\end{aligned}
\end{equation}
sourced by the three-form field strength
\begin{align}\label{3FormBB}
H_{1,txr} = H_{2,tyr} = -m \,,
\end{align}
with all other components of $H_{I,abc}$, which are not the permutation of the indices of $H_{1,txr}$ or $ H_{2,tyr}$, equal to zero. It should not come as a surprise that the background geometry is the same as in the Hodge-dualised bulk theory studied in \cite{Andrade:2013gsa}.\footnote{For other solutions in a Hodge-dualised theory with scalar fields, see e.g. \cite{Bardoux:2012aw,Gouteraux:2014hca,Andrade:2013gsa}.}

The holographic dictionary for the gravitational part of the action \eqref{eq:action1}, including the relevant Gibbons-Hawking term, the counter-terms and the asymptotic near-boundary expansions of the metric, which determine the source and the dual $T^{\mu\nu}$ expectation value, are well known. Here, we focus on the holographic dictionary for the two-form gauge fields $B_{I,ab}$ by extending the discussion of \cite{Grozdanov:2017kyl,Hofman:2017vwr}. We find that their near-boundary expansions (for the fields with the boundary spacetime components) are
\begin{equation}\label{eq:DefBfieldNB}
B_{I,\mu\nu} = r\CJ_{I,\mu\nu} + \hat B_{I,\mu\nu} + \CO(1/r) \, .
\end{equation}
By executing the procedure of holographic renormalisation (see Refs. \cite{deHaro:2000vlm,Grozdanov:2017kyl,Hofman:2017vwr}), we find that the equations of motion for the bulk gauge fields, 
\begin{equation}
\partial_a \left(\sqrt{-G} H^{a\mu\nu}_I \right) = 0 \,,
\end{equation}
directly imply the conservation of the (expectation value of the) boundary two-form currents, $\la J^{\mu\nu}_I\ra$. In terms of the bulk fields, we find that   
\begin{equation}\label{eq:DefCurrent}
\la J^{\mu\nu}_I \ra = \frac{\CH^{\mu\nu}_I }{r}  = \CJ^{\mu\nu}_I \,,
\end{equation}
where in the notation of \cite{Grozdanov:2017kyl}, $\CH_{I,\mu\nu} \equiv n^a H_{I,a\mu\nu}$ and $n^a \equiv \sqrt{G^{rr}} \delta ^{ar}$ is a unit vector normal to the boundary surface. 

The on-shell action computed from \eqref{eq:action1} also diverges in the two-form field sector and one is required to introduce the following boundary counter-terms to cancel the divergence:
\begin{align}\label{OnShellCT1}
S_{\partial M} \supset \frac{1}{4\kappa(\Lambda)} \sum_I\int_{r = \Lambda} d^3x \sqrt{-\gamma}\, \CH_{I,\mu\nu}\CH^{\mu\nu}_{I}\, ,
\end{align}
where $r = \Lambda$ is the radial position at which we place the boundary brane with the induced metric $\gamma_{ab}$---i.e. the UV cut-off---and $\kappa$ is a cut-off $\Lambda$-dependent coupling constant of which the meaning will become apparent below. After substituting the near-boundary expansions into Eq. \eqref{OnShellCT1} and using Eq. \eqref{eq:DefCurrent}, we find that the boundary counter-terms are double-trace deformations of the source terms $\sum_I \int d^3x \, J^{\mu\nu}_I b_{I,\mu\nu}$:
\begin{align}\label{eq:SbndDoubleTrace}
S_{\partial M} \supset \frac{\lambda(\Lambda)}{4} \sum_I \int d^3 x \, J_{I,\mu\nu}J^{\mu\nu}_I  \,, 
\end{align}
where $\lambda \equiv \Lambda / \kappa$ is a scale-dependent coupling constant of the boundary field theory double-trace deformations. The bulk/boundary dictionary that we derived from holographic renormalisation is therefore again analogous to the higher-dimensional situation studied in \cite{Grozdanov:2017kyl,Hofman:2017vwr} where the coupling multiplying the double-trace deformation---i.e. the boundary Maxwell action $F^2$---was the logarithmically running marginal $U(1)$ coupling (the electric charge) of the dynamical electromagnetic field which gauged the strongly coupled matter sector.

In the process of relating the boundary sources $b_{I,\mu\nu}$ to the bulk fields $B_{I,\mu\nu}$, a consistent implementation of the double-trace deformations \eqref{eq:SbndDoubleTrace} prompts us to impose mixed boundary conditions at the cut-off surface $r=\Lambda$ of the form \cite{Witten:2001ua,Berkooz:2002ug,Papadimitriou:2007sj,Grozdanov:2017kyl,Hofman:2017vwr}\footnote{A similar procedure of holographic renormalisation was also employed in the context of $AdS_2$ holography and in an analysis of subtracted geometries \cite{An:2016fzu,Cvetic:2016eiv}.}
\begin{equation}\label{eq:defSource}
b_{I,\mu\nu} = \frac{1}{2} \left( \Lambda \,\CJ_{I,\mu\nu} + \hat B_{I, \mu\nu} \right)- \frac{\Lambda}{2\kappa(\Lambda)} \CJ_{I,\mu\nu} \,,
\end{equation}
where the numerical factors are chosen in a way that keeps the prefactor of the source term $J^{\mu\nu}b_{\mu\nu}$ equal to one, which is consistent with the conventions used in \cite{Grozdanov:2016tdf,Grozdanov:2017kyl}. Now, a physical source cannot depend on the cut-off scale $\Lambda$, hence
\begin{align}\label{RGofB}
\frac{d b_{I,\mu\nu}}{d\Lambda} = 0 \,,
\end{align} 
which gives rise to a renormalisation group (RG) equation for a running double-trace coupling
\begin{align}\label{eq:betaFunc}
\frac{d \lambda(\Lambda)}{d\Lambda} = \frac{d}{d\Lambda}\left(\frac{\Lambda}{\kappa(\Lambda)}\right) = 1 \,.
\end{align}
The solution of this beta function equation is a linearly running
\begin{align}\label{Runninglambda}
\lambda(\Lambda) = \frac{\Lambda}{\kappa(\Lambda)} = \Lambda - \CM \,,
\end{align}
where $\CM$ is an integration constant of the beta function equation \eqref{eq:betaFunc} with a mass dimension one. This is the renormalisation group scale, of which the value needs to be imposed by using external experimental input, in the same way that the renormalised electromagnetic coupling had to be fixed in \cite{Grozdanov:2017kyl}. In the holographic on-shell action, the $\Lambda$-dependent linear divergence now disappears as a result of the cancellation between the on-shell part of the action coming from \eqref{eq:action1} and the counter-terms \eqref{eq:SbndDoubleTrace}. The renomalised double-trace deformation term is thus finite in the limit of $\Lambda\to\infty$:
\begin{align}
S_{\partial M} &\supset - \frac{\CM}{4} \sum_I \int d^3 x \, J_{I,\mu\nu}J^{\mu\nu}_I \\ 
&= - \frac{\CM}{2} \sum_I \int d^3 x \, \partial_\mu \psi_I \partial^\mu \psi_I \, , \label{DoubleTraceScalar}
\end{align}
where in the last line, we used the Hodge dualised, scalar field representation of the conserved two-forms discussed around Eqs. \eqref{ScalarHodge1} and \eqref{ScalarHodge2}, i.e. $\star\, J_I = \xi_I = d \psi_I $. The action in Eq. \eqref{DoubleTraceScalar}, which arose as a direct consequence of holographic renormalisation, has the form of a kinetic term of two dimensionless scalar fields with $\lambda$ from Eq. \eqref{eq:SbndDoubleTrace} a relevant coupling constant with a mass dimension one. Thus, the expression for the on-shell action is consistent with the way we wrote $J_I$ in terms of closed and exact forms in Section \ref{sec:Introduction}. Finally, we note that from the point of view of elasticity theory, the value of the renormalised coupling $\lambda$, or the scale $\CM$, is related to the size of one of the elastic moduli \cite{PhysRevA.6.2401,Delacretaz:2017zxd}. A material-specific measurement of the elastic modulus in the IR state of the theory can therefore set the value of the renormalised microscopic coupling constant $\lambda$ when the relation between them is known.

\subsection{Thermodynamics and hydrodynamics}\label{subsec:thermo}

As the first step in the analysis of the IR properties of the dual boundary field theory, we study its thermodynamic quantities and the equation of state that follow from the action \eqref{eq:action1} and the solution in Eqs. \eqref{eq:DefMetric} and \eqref{3FormBB}. First, we compute the Euclidean on-shell action to find the pressure (see e.g. \cite{Grozdanov:2017kyl}). The entropy density and temperature can be extracted from the horizon limit of the metric alone. Together, 
\begin{equation}
  \begin{aligned}
  p &= r_h^3\left( 1 + \left(\bar \CM-\frac{3}{2} \right)\bar m^2 \right),\\
  s &= 4\pi r_h^2 ,\\
  T &= \frac{r_h}{4\pi} \left( 3-\frac{1}{2}\bar m^2  \right),
\end{aligned}
\end{equation}
where we have introduced two dimensionless quantities $\bar \CM \equiv \CM/r_h$ and $\bar m \equiv m/r_h$, with
\begin{align}
r_h = \frac{1}{6} \left( 4\pi T + \sqrt{ (4\pi T)^2 + 6 \rho^2  } \right) \,.  
\end{align}
By computing the equilibrium holographic stress-energy tensor, one can obtain all three of its diagonal components, 
\begin{align}
\la T^{tt}\ra &= \varepsilon = r_h^3\left(2+\left(\bar \CM-1\right)\bar m^2  \right),\\
\la T^{xx}\ra &= \la T^{yy}\ra = p - \mu\rho = r_h^3\left(1-\frac{\bar m^2}{2}  \right) \,. \label{EffPressureEq}
\end{align}
Requiring that the temperature remains non-negative constrains the parameter $\bar m$ to $\bar m \in [0, \sqrt{6}]$. We further note that the following two regimes, $\bar m \in [0, \sqrt{2} )$ and $\bar m \in (\sqrt{2}, \sqrt{6}]$, result in the pressure shifted by the tension of the flux lines to be either positive or negative, respectively (cf. Eq. \eqref{EffPressureEq}). 

Now, consider the two-form gauge field sector from which the number density of the topological defects is found to be
\begin{align}\label{RhoM}
\rho = \la J^{tx}_{1} \ra = \la J^{ty}_{2}\ra = m \,,
\end{align}
and the corresponding chemical potential can be read off from the source, i.e. $\mu_{I} =b_{tI}-b_{It}$, with $b_{\mu\nu}$ defined in Eq. \eqref{eq:defSource}. In this case, 
\begin{align}\label{ChemPot}
\mu = \mu_{1} = \mu_{2} =  (\bar\CM-1)\,m r_h \,.
\end{align}
For the chemical potential to be positive, it is important that the renormalisation group scale be $\bar\CM >1$. Note that one can easily check that the above thermodynamic quantities satisfy Eq. \eqref{Thermo1}, i.e.
\begin{align}
\varepsilon + p -s\, T = \sum_I \mu_{I}\rho_{I} = 2\mu\rho \,,
\end{align}
and also the required relation between $\la T^{xx}\ra$ and pressure,
\begin{align}
p-\la T^{xx}\ra =  \mu\rho \,,
\end{align}
as dictated by Eq. \eqref{eq:DefIdealStessEn}.

By using the holographic equation of state and hydrodynamic (field theory) results from Section \ref{sec:EFT}, one can predict the speed of sound in the transverse channel to be 
\begin{equation}
\begin{aligned}
\CV_A^2 &= \frac{1}{3} \left(\frac{(\bar \CM-1)\bar m^2}{1+ \frac{\bar m^2}{3}\left(\bar \CM-\frac{3}{2} \right)}\right),\\
&\approx  \frac{1}{3} (\bar \CM-1)\bar m^2 + \CO(\bar m^4) \,,
\end{aligned}
\label{eq:VAinAdS}
\end{equation}
where the second line is the leading-order expansion in small density compared to the temperature. In Section \ref{sec:finitewall}, we will show analytically to $\CO(m^2)$ that the speed of the propagating shear mode, computed directly from a holographic result for a transverse two-point function, indeed obeys Eq. \eqref{eq:VAinAdS}. Finally, as we take the limit of $\bar\CM \to \infty$, keeping $\bar m$ fixed, which in terms of the field theory observables is the limit of $T / \CM \to 0$, the speed of the transverse sound mode tends to the speed of light, $\CV_A^2 = 1$. The importance of keeping the chemical potential \eqref{ChemPot} positive is apparent from Eq. \eqref{eq:VAinAdS}; the transverse sound mode would otherwise become unstable. A possible scenario that could elucidate the nature of this instability in one in which there exists another stable phase of matter for $\CM < 1$ in the $(T,\rho,\CM)$ phase diagram. Such a phase would correspond to an independent, stable branch of gravitational backgrounds. However, establishing the existence of such a phase may require one to employ numerical methods to solve the bulk system of partial differential equations, which is beyond the scope of this work. For this reason, we postpone a detailed analysis of the dual boundary phase diagram to future works.

\subsection{The spectrum at zero density, $\rho=0$}\label{sec:zerowall}

As discussed in Section \ref{sec:Introduction}, at a vanishing equilibrium number density of line defects, $\rho = 0$ (or by Eq. \eqref{RhoM}, $m=0$), one expects the IR limit of the system to be controlled by the fluid and not the elastic component of the thermal state, i.e. by the purely diffusive modes in the retarded two-point function spectra of $\la T^{\mu\nu}T^{\rho\sigma}\ra_R$ and $\la J^{\mu\nu}J^{\rho\sigma}\ra_R$. To show that this is true in our holographic setup, we perturb the bulk fields to first order, $G_{ab} \to G_{ab} + \delta G_{ab}$, $B_{I,ab} \to B_{I,ab} + r_h^2 \delta B_{I, ab}$, write the perturbations in terms of their (boundary spacetime) Fourier decomposition, i.e. as $e^{-i\omega t + i k x} \delta G_{ab}(\omega,k,u)$ and similarly for $\delta B_{I,ab}$, where we have introduced a new radial coordinate $u = r_h/r$. In the radial gauge, and in the sector of transverse fluctuations that are odd under $y\to -y$, we find that the components of $\delta B_{2,ab}$ decouple and can be set to zero. Thus, the case studied here is precisely equivalent to the transverse channel in a $\rho=0$ theory with a single $U(1)$ generalised global symmetry. The remaining fluctuations can be arranged into two independent gauge-invariant combinations
\begin{equation}\label{GaugeInvModes}
\begin{aligned}
Z_- &= \omega \, \delta G^{y}_{~x} + k \,\delta G^{y}_{~t} \,,\\
Z_+ &= \omega \, \delta B_{1,xy} + k \, \delta B_{1,ty} \,.
\end{aligned}
\end{equation}

The dynamical equation for $Z_-$ and $Z_+$ decouple. The lowest-lying IR mode coming from $Z_-$, which determines the hydrodynamic pole of the retarded transverse $G^{\mu\nu,\rho\sigma}_{TT,R}(\omega,k)$ correlators, is the usual momentum diffusion pole of a thermal M2 brane that follows from the standard (Dirichlet boundary condition) quasinormal mode equation $Z_- = 0$ imposed at the boundary. To leading order, its dispersion relation is \cite{Herzog:2002fn}
\begin{align}\label{M2diffusion}
\omega = -i \frac{1}{4\pi T} k^2 \,.
\end{align}

On the other hand, the dynamics of the gauge-invariant combination of the two-form gauge fields, which determine the spectrum of the retarded transverse $\la J^{\mu\nu}_1J_1^{\rho\sigma}\ra_R$ correlators, obeys the following differential equation:
\begin{align}\label{eq:Z2zeroden}
Z_+'' + \left(\frac{2}{u} + \frac{\omega^2 f'}{f(\omega^2-k^2f)}  \right)Z_+' + \left(\frac{\omega^2-k^2f}{r_h^2f^2}  \right)Z_+ = 0 \,. 
\end{align}
At non-zero temperature, this equation can be solved in a hydrodynamic expansion with $|\omega|/T \ll1$ and $|k|/T\ll1$. To find the diffusive pole, it is sufficient to scale $\omega/T \sim (k/T)^2$. Then, to first order, the solution satisfying in-falling boundary conditions at the horizon is
\begin{equation}\label{eq:solZp}
Z_+ = \CC_+ f(u)^{-\frac{i\omega}{4\pi T}} \left[1 + \frac{ik^2}{r_h \omega}\left(\frac{1}{u}-1  \right) + \ldots \right] \,.
\end{equation}
It is now essential to note that because we are working with mixed boundary conditions in a boundary theory with double-trace deformations, the quasinormal mode prescription of setting $Z_+ = 0$ no longer determines the poles of the retarded $\la J^{\mu\nu}J^{\rho\sigma}\ra_R$ correlators. The double-trace deformation shifts the pole of the correlator. The correct procedure (a modified quasinormal mode prescription) in this case is to determine to sources $b_{I,\mu\nu}$ from the definition in Eq. \eqref{eq:defSource} and demand that they vanish: 
\begin{align}\label{QNMMod}
b_{I ,\mu\nu} = 0 \,.
\end{align}
In this example, we evaluate the solution $Z_+$ from Eq. \eqref{eq:solZp} at the cut-off surface $u = r_h / \Lambda$ and expand it around $\Lambda \to \infty$:
\begin{align}\label{Zexp1}
Z_+ = \CC_+ \left[ \frac{i k^2}{r_h^2 \omega} \Lambda + \left(1 - \frac{i k^2}{r_h \omega} \right) + \ldots    \right] \,.
\end{align}
Then, using the gauge-invariant definition of $Z_+$ from Eq. \eqref{GaugeInvModes}, mixed boundary conditions \eqref{eq:defSource} and the running coupling \eqref{Runninglambda}, we can write
\begin{align}\label{Zexp2}
Z_+ = 2 Z_+^{(0)} + \left(\Lambda - \CM \right) Z_+^{(1)} + \ldots \,,
\end{align}
where $Z_+^{(0)}$ is the source and $Z_+^{(1)}$ is related to the dual expectation value of a gauge-invariant combination of $J^{\mu\nu}_1$ in the transverse channel. Comparing Eqs. \eqref{Zexp1} and \eqref{Zexp2}, we find
\begin{align}\label{Zpm}
Z_+^{(0)} = \CC_+ \frac{i k^2 \left(\bar\CM - 1\right) + r_h \omega}{2 r_h \omega}\,, && Z_+^{(1)} = \CC_+ \frac{i k^2}{r_h^2 \omega} \,.
\end{align}
The condition \eqref{QNMMod}, which determines the poles of the correlators, now requires us to set 
\begin{align}\label{QNMModZ}
Z_+^{(0)} (\CM)= \lim_{\Lambda\to\infty} \frac{1}{2} \left[ Z_+(\Lambda) - \lambda(\Lambda,\CM) \frac{\partial Z_+ (\Lambda)}{\partial \Lambda}    \right] = 0 \,,
\end{align} 
and look for the dispersion relations $\omega(k)$ that satisfy this equation. It is important to note that because of the renormalisation group equation \eqref{RGofB}, the prescription \eqref{QNMMod} (or its gauge-invariant equivalent in \eqref{QNMModZ}) implies that the poles do not depend on the unphysical cut-off $\Lambda$, only the renormalisation group scale $\CM$ (see also \cite{Grozdanov:2017kyl,Hofman:2017vwr}). At the leading order in the hydrodynamic gradient expansion (cf. Eq. \eqref{Zpm}), we then find a diffusive dispersion relation
\begin{equation}\label{eq:LatticeDiffu}
\omega = -i \frac{\left( \bar \CM - 1  \right)}{r_h} k^2 = - i \frac{3}{4\pi T} \left( \frac{3 \CM}{4\pi T}  - 1\right) k^2 \,,
\end{equation}
where the diffusion constant now depends on the renormalisation group scale $\CM$ as in the higher-dimensional analogue studied in \cite{Grozdanov:2017kyl,Hofman:2017vwr}. In those works, the diffusion constant (but not the resistivity) depended on the renormalised electric charge (or the energy scale of the Landau pole). Furthermore, as discussed in \cite{Grozdanov:2017kyl}, the system can become unstable for certain parameter regimes of $\CM$. In this case, we see that at the least, we need to require that $\bar\CM > 1$ or $T/\CM < 3 / 4\pi$---the temperature scale needs to be below the renormalisation group scale. Moreover, for \eqref{eq:LatticeDiffu} to be a valid solution within the hydrodynamic regime, we also require $\CM k^2 / T^3 \ll 1$.

Beyond the hydrodynamic limit, the spectrum of the $\la J^{\mu\nu}_I J^{\rho\sigma}_I\ra_R$ correlators at zero density also depends on temperature $T$ and the renormalisation group scale $\CM$. At energy scales above the extreme IR limit, the spectrum contains an infinite set of higher quasinormal modes, which can readily be found in holography. A commonly observed phenomenon in spectra that depend on several scales is the destruction of hydrodynamics in the parameter regime where the scale separation between hydrodynamics and higher-energy modes ceases to exist. For example, a coupling constant dependent spectrum at intermediate coupling exhibits modes with purely imaginary, gapped dispersion relations \cite{Grozdanov:2016fkt,Grozdanov:2016vgg}, of which the leading mode limits the regime of validity of hydrodynamics at intermediate coupling \cite{Grozdanov:2016fkt,Grozdanov:2016vgg,Grozdanov:2016zjj,Andrade:2016rln,DiNunno:2017obv}. In the transverse shear channel, at a fixed momentum and as the field theory coupling is tuned from infinity towards weak coupling, the hydrodynamic diffusive mode collides with the non-hydrodynamic, gapped mode on the imaginary $\omega$ axis at some critical value of the coupling. Similarly, at a fixed intermediate coupling and an increasing $k/T$, the collision occurs at some critical momentum $k_c/T$. After the collision, the two modes move off the axis and $\omega(k)$ acquire a real part \cite{Grozdanov:2016fkt,Grozdanov:2016vgg}. This phenomenon was also observed in models closely related to this work \cite{Davison:2014lua,Alberte:2017cch} as well as in the holographic dual of MHD with a generalised global symmetry \cite{Hofman:2017vwr}. In our theory, we observe precisely the same dynamics of the poles, which we will now analyse both numerically and analytically. 

\begin{figure*}[t]
\begin{center}
\includegraphics[width=0.45\textwidth]{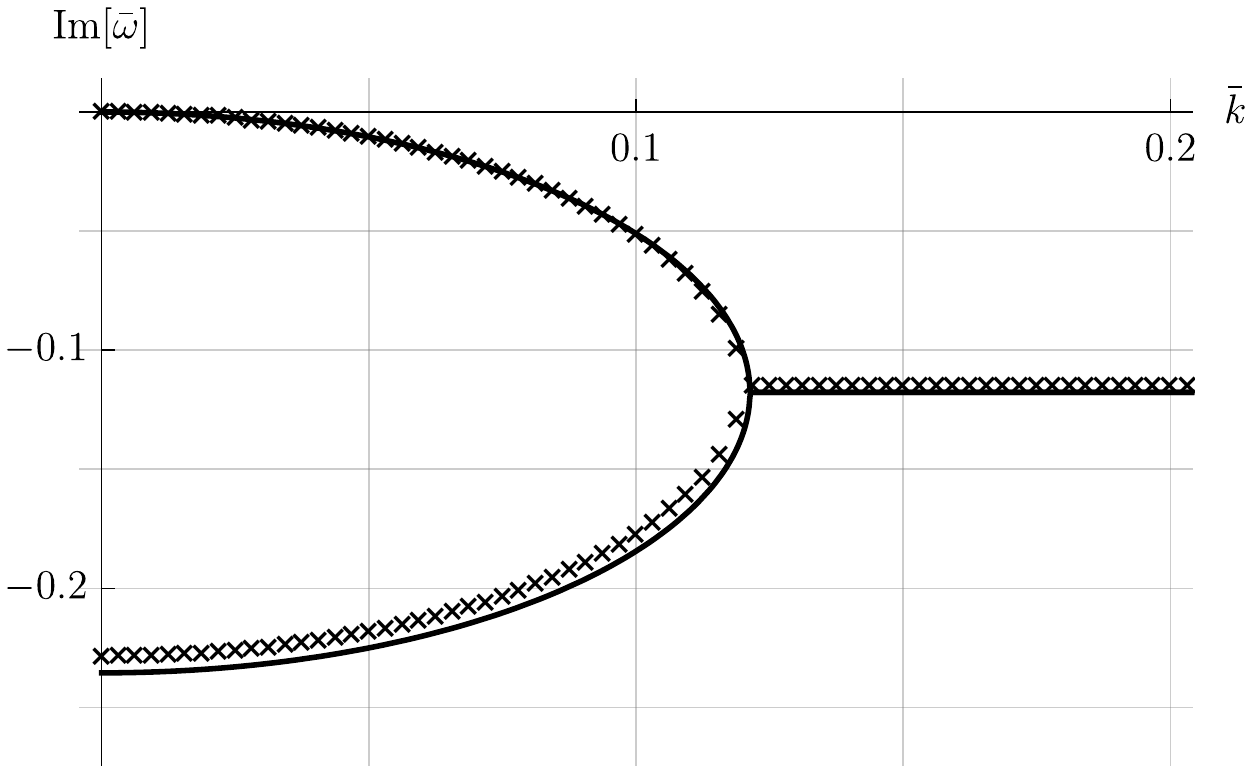} 
\hspace{0.05\textwidth}
\includegraphics[width=0.45\textwidth]{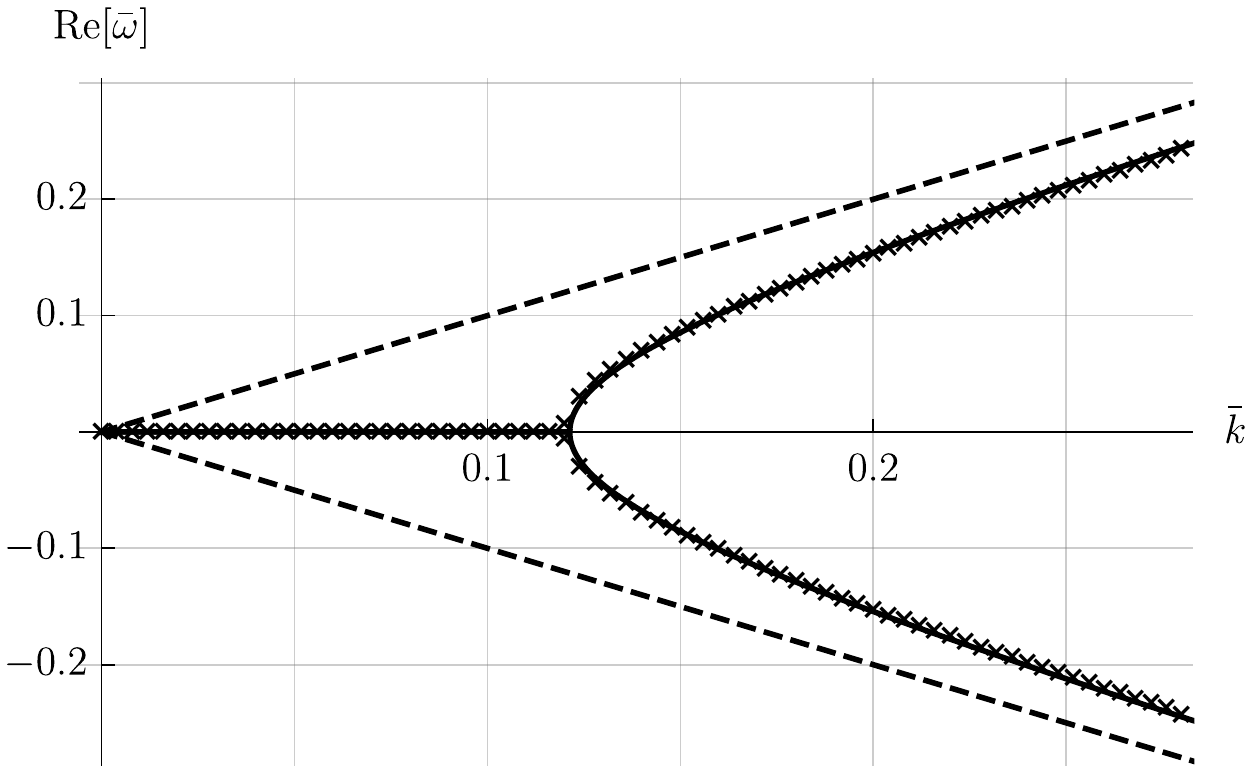}
\end{center}
\caption{Plots of imaginary and real parts of the hydrodynamic and the non-hydrodynamic dispersion relations at $\bar\CM = 5$, with $\bar\omega \equiv \omega / r_h$ and $\bar k \equiv k / r_h$. Crosses depict the numerically computed poles of the retarded transverse part of the $\la J^{\mu\nu}_1 J^{\rho\sigma}_1\ra_R$ correlator at zero density, which follow from the prescription \eqref{QNMModZ}. The solid lines show the analytical approximation \eqref{Full2solutions} to the dispersion relations. The dashed lines represent the linear dispersion relation $\bar\omega = \pm \bar k $.}
\label{fig:M5}
\end{figure*}

The dispersion relation of the diffusive mode in \eqref{eq:LatticeDiffu} makes it clear that the dimensionless frequency $|\omega| / T$, where we think of $T$ as the cut-off scale of the effective hydrodynamic theory, can move outside of the hydrodynamic regime for large $\CM/T$, large $k/T$ or large $\CM k^2/T^3$. These are precisely the parameter regimes in which the (leading) non-hydrodynamic, gapped mode with a purely imaginary frequency has $|\omega|$ that is comparable to $|\omega|$ of the diffusive mode. As in \cite{Grozdanov:2016fkt,Grozdanov:2016vgg}, the two modes collide, move off the imaginary $\omega$ axis, become propagating and at large momentum travel with the speed of light. As shown in \cite{Grozdanov:2016fkt,Grozdanov:2016vgg}, this behaviour can be understood analytically in the double-scaling parameter regime where the collision, controlled by an independent scale, occurs in the (hydrodynamic) regime of small $|\omega|/T$ and $k/T$. In the present case, following \cite{Grozdanov:2016fkt,Grozdanov:2016vgg}, we expand $Z_+$ to sub-leading order in $|\omega|/T \sim k/T \ll 1$ and keep $T / \CM \ll 1$ (or $\bar\CM \gg 1$). The polynomial equation, which determines the dual dispersion relations, again follows from setting $Z_+^{(0)} = 0$ (cf. Eq. \eqref{QNMModZ}):
\begin{align}
\left(1 - \frac{\omega}{\omega_{\mathfrak g}}  \right) \omega + i  \frac{\left(\bar\CM-1\right)}{r_h} k^2  =  0 \,,
\label{eq:NonHydroSpectrum}
\end{align}
where
\begin{equation}
\omega_{\mathfrak g} \equiv - \frac{i r_h }{ \bar \CM-1 +\frac{1}{2}\left(\ln 3 - \frac{\pi}{3\sqrt{3}} \right) } \,.
\end{equation}
The quadratic equation \eqref{eq:NonHydroSpectrum} has the following two solutions \cite{Grozdanov:2016fkt,Grozdanov:2016vgg}:
\begin{align}\label{Full2solutions}
\omega_\pm = \frac{\omega_{\mathfrak{g}}}{2} \left( 1 \pm \sqrt{1 - \frac{4\left(\bar\CM-1\right) k^2}{r_h \left|\omega_{\mathfrak g} \right|  } } \right) \,,
\end{align}
which can be expanded for $k / r_h \sim k/T \ll 1$ to give
\begin{align}
\omega_- &= - i \frac{\left(\bar\CM-1\right)}{r_h} k^2  + \ldots \,, \label{Full2solutionsHydro} \\
\omega_+ &= \omega_\mathfrak{g} + i \frac{\left(\bar\CM-1\right)}{r_h} k^2 +\ldots \, . \label{Full2solutionsNonHydro} 
\end{align}
Thus, \eqref{Full2solutionsHydro} is the diffusive mode from Eq. \eqref{eq:LatticeDiffu} and \eqref{Full2solutionsNonHydro} is the new, non-hydrodynamic gapped mode. It is clear that the latter solution is only reliable when $|\omega_{\mathfrak g}| / T \ll 1$, i.e. when $1/\bar\CM \sim T / \CM \ll 1$. Furthermore, using Eq. \eqref{Full2solutions} and the condition $\text{Im}[\omega_-(k_c)] = \text{Im}[\omega_+(k_c)]$, we can determine the (analytical approximation to the) critical momentum $k_c$ at which the two modes collide (see \cite{Grozdanov:2016fkt,Grozdanov:2016vgg}):
\begin{align}
k_c = \frac{1}{2} \sqrt{ \frac{ r_h \left| \omega_{\mathfrak g} \right|  }{\bar\CM - 1 } } \,.
\end{align}
Therefore, in the limit of large $\bar\CM$, the collision occurs both at small $|\omega |/ T$ and $k/T$, i.e. $\omega_\pm (k_c) / r_h \approx 1 / (2\bar\CM) $ and $k_c / r_h \approx 1 / (2\bar\CM)$, which is within the regime assumed in the derivation of the dispersion relations \eqref{Full2solutions}.

\begin{figure}[h]
\begin{center}
\includegraphics[width=0.9\textwidth]{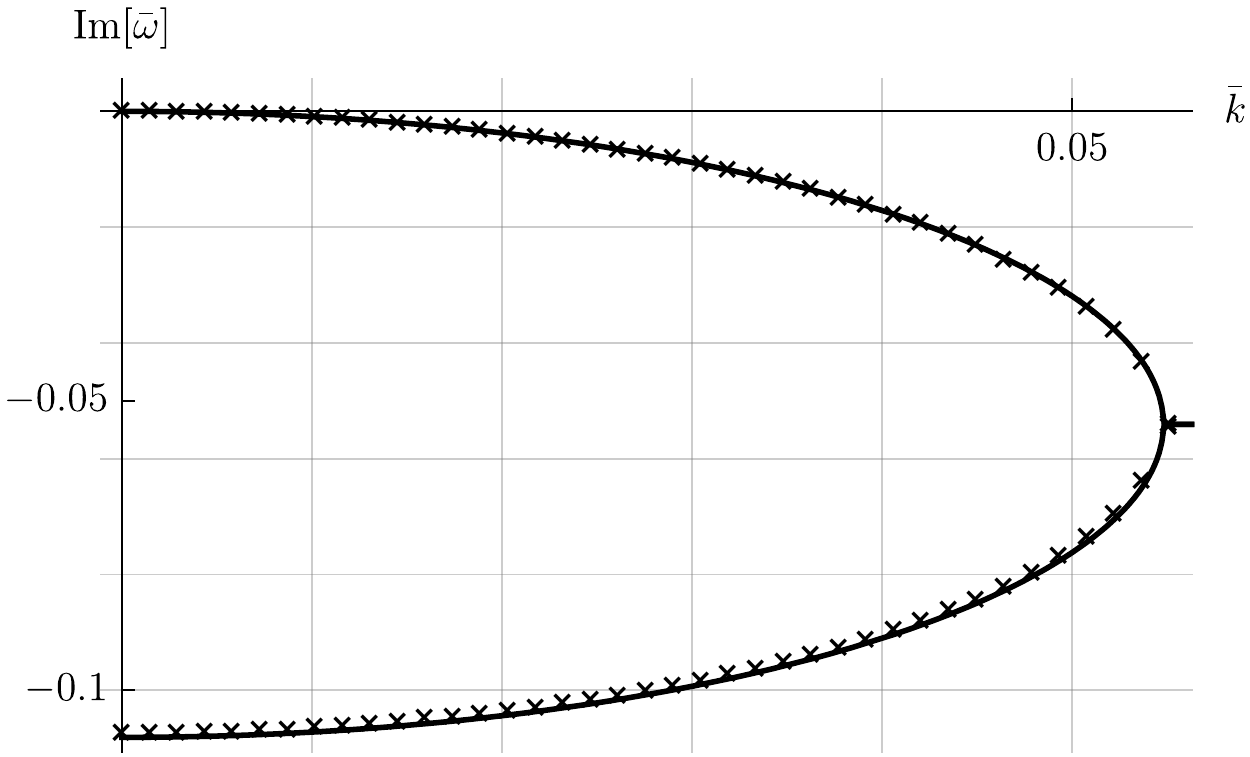} 
\end{center}
\caption{Imaginary parts of the hydrodynamic and the non-hydrodynamic dispersion relations of the retarded transverse $\la J^{\mu\nu}_1 J^{\rho\sigma}_1\ra_R$ correlator at zero density, plotted at $\bar\CM = 10$ up to the point of the collision. Crosses depict the full numerically obtained poles and the solid lines are their analytical approximations from Eq. \eqref{Full2solutions}.}
\label{fig:M10}
\end{figure}

It is rather interesting to note that in the limit of $k / T \to \infty$, while keeping $\CM/T$ fixed, the analytical dispersion relations \eqref{Full2solutions} give
\begin{align}
\omega_{\pm} = \pm   \sqrt { \frac{ \left|\omega_{\mathfrak g} \right| \left(\bar\CM - 1 \right) }{r_h} }  k  - \frac{ i \left|\omega_{\mathfrak g} \right| }{2 r_h} + \ldots \, , 
\end{align}
which, furthermore, in the limit of $\bar\CM \sim \CM /T \to \infty$ reproduce a dispersion relation of waves travelling with the speed of light,
\begin{align}\label{DispLight}
\omega_\pm = \pm k \, .
\end{align} 
Finally, note that in the limit of $\bar\CM \to \infty$ (keeping $k/T$ fixed), \eqref{Full2solutions} also tends to \eqref{DispLight}. Thus, even though Eq. \eqref{Full2solutions} was derived by assuming a hydrodynamic expansion, in the regime of $T / \CM \ll 1$, $\omega_\pm$ correctly reproduce the UV behaviour of the modes. The comparison between numerical and analytical results is depicted in Fig. \ref{fig:M5} for the value of the renormalisation group scale set to $\bar\CM = 5 $. Although the analytics capture the qualitative behaviour of the two poles for all $\CM$, as is clear from the above discussion, the approximation improves as $\bar\CM$ is tuned to be larger. While the agreements between analytics and numerics is already nearly indistinguishable to the eye for $\text{Re}[\omega]$ at $\bar\CM = 5$, the improvement for the imaginary part can be seen from a comparison of Figs. \ref{fig:M5} and \ref{fig:M10}.

\subsection{The spectrum of the transverse $\la J^{\mu\nu}_1J_1^{\rho\sigma}\ra_R$ at zero density and zero temperature, $\rho = 0$ and $T = 0$}\label{sec:zerowallzerotem}

In the previous section, massless modes, which propagate at the speed of light were seen to emerge in the spectrum of $\la J^{\mu\nu}_1J_1^{\rho\sigma}\ra_R$ from a pair of a hydrodynamic and a gapped, non-hydrodynamic mode in the limit of $T\to0$. Here, with the help of restored Lorentz invariance at $T=0$, we analyse the full correlator $\la J^{\mu\nu}_1J_1^{\rho\sigma}\ra_R$, which confirms the existence of massless modes in the spectrum. Beyond that, we show that the spectrum also contains a pair of massive modes of which the mass is set by the (dynamically generated) renormalisation group scale $\CM$, which further displays the importance of a careful renormalisation group analysis of our holographic setup. 

We begin by writing down a tensorial decomposition of the Wick-rotated, Euclidean two-point correlation function $G^{\mu\nu,\rho\sigma}_{J_1J_1}(k)$. Using the Ward identity and antisymmetric property of $J^{\mu\nu}$, we can write \cite{Kovtun:2005ev,Hofman:2017vwr}
\begin{align}\label{GinEucl}
G^{\mu\nu,\rho\sigma}_{J_1J_1}(k) = \left[P^{\mu\nu\rho\sigma} + \frac{1}{k^2}Q^{\mu\nu\rho\sigma}\right]F(|k|,\CM_E) \,,
\end{align}
where $k^2 \equiv k_\mu k^\mu$, $|k| = \sqrt{k_\mu k^\mu}$ and $k^\mu=(\omega_E,\vec{k})$. The renormalisation group scale $\CM_E$ is an imaginary energy scale in the space of Wick-rotated $\omega_E$. The renormalisation group condition \eqref{QNMModZ} must now be chosen with more care---i.e. we must make sure that we correctly analytically continue $\omega^2/\CM^2$ to $\omega_E^2/\CM_E^2$. We present details on how this procedure should be executed in Appendix \ref{appendix:AnalyticCont}. Here, we only note that we directly use $\CM_E$ in the place of $\CM$ in Eq. \eqref{QNMModZ} and in continuing the correlator to real time, set $\CM_E = i  \CM$.\footnote{We note that in quantum field theory, the question of finding a relativistically invariant regulator and renormalising a theory with real time is a difficult problem. We refer the reader to discussions in \cite{Polonyi:2001se,Floerchinger:2011sc,Grozdanov:2011aa,Polonyi:2017xdb}.} The projectors $P^{\mu\nu\rho\sigma}$ and $Q^{\mu\nu\rho\sigma}$ used in Eq. \eqref{GinEucl} are defined as 
\begin{equation}
\begin{aligned}
P^{\mu\nu\rho\sigma} &=  g^{\mu\rho}g^{\nu\sigma} - g^{\mu\sigma}g^{\nu\rho} \, ,\\
Q^{\mu\nu\rho\sigma} &= k^\mu k^\rho g^{\nu\sigma} + k^\nu k^\sigma g^{\mu\rho} - [\mu\leftrightarrow \nu, \rho\leftrightarrow \sigma] \,.
\end{aligned}
\end{equation}
This structure guarantees that $k_\mu \la J^{\mu\nu}_1J_1^{\rho\sigma}\ra = 0 $. For a system with momentum pointing along the $x$-direction, $\vec{k} = (k_x ,0)$, the transverse two-point functions are
\begin{equation}\label{3GsEuclidean}
\begin{aligned}
G^{\tau y,\tau y}_{J_1J_1}(\omega_E,k_x) &= \frac{k_x^2}{\omega_E^2+k_x^2}F(|k|,\CM_E) \,,\\
G^{xy,xy}_{J_1J_1}(\omega_E,k_x) &= \frac{\omega_E^2}{\omega_E^2+k_x^2}F(|k|,\CM_E)\,,\\
G^{\tau y,xy}_{J_1J_1}(\omega_E,k_x) &= -\frac{\omega_E\, k_x}{\omega_E^2+k_x^2}F(|k|,\CM_E) \,,
\end{aligned}
\end{equation}
where $\tau$ is the imaginary time. This reduces the problem to finding a single function $F(|k|,\CM_E)$. At $\rho=0$ and $T=0$, we can use the Euclidean $SO(3)$ symmetry to set $\omega_E =0$ and compute $F(|k|,\CM_E) = F(k_x,\CM_E)$. In this case, on the bulk side of the holographic duality, the gauge-invariant mode $Z_+$ reduces to $Z_+ = k_x \, \delta B_{1.\tau y}$ and 
\begin{align}
G^{\tau y,\tau y}_{J_1J_1}(0,k_x) = F(k_x,\CM_E) \, . 
\end{align}
To compute the transverse Euclidean $G^{\mu\nu,\rho\sigma}_{J_1J_1}(0,k_x)$ from holography, one can thus take the zero temperature limit of \eqref{eq:Z2zeroden}, set $\omega =0$ and compute $G^{\tau y,\tau y}_{J_1J_1}(0,|k|)$, which determines $F(|k|,\CM_E)$. $F(|k|,\CM_E)$ then fixes all three correlators via Eq. \eqref{3GsEuclidean}. In terms of the radial coordinate $r$, we find the following solution for $Z_+$ that is non-singular at the horizon: 
\begin{equation}
Z_+ = k_x\, \delta B_{\tau y} = \CC_+ r e^{-k_x/r} \, .
\end{equation}
Using the prescription for computing correlation functions, which we discussed in Section \ref{sec:HolRG}, we can show that 
\begin{equation}\label{F}
F(k_x,\CM_E) \propto \frac{1}{\CM_E \left( 1-k_x/\CM_E \right)} \,.
\end{equation}
Thus, we can conclude from \eqref{3GsEuclidean} and \eqref{F} that the transverse Euclidean correlation functions of $J^{\mu\nu}_1$ are governed by two sets of modes, $\omega_E^2 + k_x^2 = \CM_E^2 $ and $\omega_E^2 + k_x^2 = 0$. By analytically continuing back to real time, these two (pairs of) poles give rise to retarded correlators with a massless mode that travels at the speed of light and a gapped mode of which the gap (the mass scale) is set by the renormalisation scale $\CM$:
\begin{align}
\omega &= \pm k_x \, ,\\
\omega &= \pm \sqrt{\CM^2 + k_x^2 } \, .
\end{align}
The first of the two modes was recovered in Section \ref{sec:zerowall} by the hydrodynamic analysis of the thermal spectrum in the $T \to 0$ limit. The massive gapped mode lies outside the regime of validity of our analytical calculation in Section \ref{sec:zerowall}.

\subsection{The spectrum at non-zero density, $\rho >0$}\label{sec:finitewall} 

In the final part of our holographic investigation, we turn our attention to a state with a non-zero density of perpendicular line defects, $\rho > 0$, and non-zero temperature, $T>0$. In this case, the two gauge-invariant bulk modes corresponding to gravitational and two-form gauge field perturbations, $Z_-$ and $Z_+$ (cf. Eq. \eqref{GaugeInvModes}), are coupled and satisfy the following set of differential equations:
\begin{align}
&Z_\pm'' \pm  \left[ \frac{2}{u} \pm \frac{\omega^2f'}{f(\omega^2-k^2f)}\right] Z'_\pm + \left[ \frac{\omega^2-(k^2+m^2)f}{r_h^2 f^2} \right] Z_\pm\nn
& - \left[ \frac{\omega k m  f'}{r_h u^{\pm2} f (\omega^2-k^2f)} \right] Z_\mp  = 0 \, . \label{EqsFull}
\end{align}
For our particular choice of momentum, the transverse fluctuations of $\delta B_{2,\mu\nu}$ remain decoupled and can be consistently set to zero. 

We first look for the analytical solutions to \eqref{EqsFull} in the hydrodynamic expansion with $|\omega| / T \ll 1$, $k / T \ll1$ as well as in the limit of a small density, $ \rho / T = m /T \ll 1$. For the expansion, we use the following small parameter scaling: $\bar \omega \sim \bar k \sim \bar m \sim \delta \ll 1$, where $\bar\omega \equiv \omega / r_h$, $\bar k \equiv k/ r_h$ and $\bar m \equiv m / r_h$. The solutions can be found by using the same procedure as in Section \ref{sec:zerowall}, only now for a coupled set of two differential equation. For present purposes, it is sufficient to find $Z_\pm$ to sub-leading order in the small parameter expansion, of which the forms will for conciseness only be stated schematically:
\begin{align}
Z_\pm = \CC_\pm f(u)^{-\frac{i\omega}{4\pi T}} &\left[ 1 + z_{1,\pm} (u,\omega, k, r_h, m) \, \delta \right. \nn
&\left. \,\,+ \,\CO(\delta^2) \right] \,.\label{eq:solHydroLim2}
\end{align} 
Once the solutions are known, then in terms of the $u$ coordinate, the (modified) quasinormal mode prescriptions for finding the dual poles of retarded correlators are (cf. Sections \ref{sec:HolRG} and \ref{sec:zerowall})
\begin{align}
\lim_{u\to 0} \left[Z_+ + u\left(1 - u \bar\CM \right) \frac{\partial Z_+}{\partial u } \right] = 0 \, , \label{Z+PrescU}\\
\lim_{u\to 0} Z_- = 0 \,,\label{Z-PrescU}
\end{align}
where \eqref{Z+PrescU} follows directly from \eqref{QNMModZ}. First, we find that Eq. \eqref{Z+PrescU} implies
\begin{align}
\CC_+ =\CC_- \left( \frac{3 r_h \omega + i k^2}{ m k}\right) \,,
\end{align}
which can be substituted into the solutions in Eq. \eqref{eq:solHydroLim2}. Finally, using the mixed boundary conditions from Eq. \eqref{Z-PrescU}, we obtain a cubic equation in $\omega$:
\begin{align}
&\omega^3 - \left( \omega_{\mathfrak g} - \frac{i k^2}{3r_h}  \right) \omega^2 -  \frac{i k^2 \omega_{\mathfrak{g}}}{3 r_h} \left(3 \bar\CM - 2 \right) \omega \nn
& + \frac{k^2 \omega_{\mathfrak{g}}}{3r_h^2} \left(m^2 + k^2 \right) \left(\bar\CM - 1 \right) = 0 \,, \label{PolyFiniteRho}
\end{align} 
which can be solved to find the dispersion relations of the three IR modes in the spectrum at $\rho > 0$, valid to sub-leading order in the small $\delta$ expansion of $Z_\pm$. In an expansion of $\bar\omega $ around a small $\bar k$, we find two propagating sound modes and a gapped mode:
\begin{align}
\omega_{1,2} =& \pm \left[ \frac{1}{3} \left( \bar\CM - 1 \right)\right]^{1/2} \bar{m} k   \nn
& - i \left( \frac{3\bar\CM - 2}{6r_h} +\ldots   \right) k^2 + \CO(k^3) \, ,\\
\omega_3 =&\,\, \omega_{\mathfrak{g}} + i \left( \frac{\bar\CM - 1 }{r_h} + \ldots \right) k^2 +  \CO( k^4 ) \, , 
\end{align}
where the ellipses denote sub-leading $m$-dependent corrections. Hence, the speed of sound computed from the holographic spectrum of two-point functions via the quasinormal mode analysis, to leading-order in $\bar m$, precisely reproduces the leading-order result from Eq. \eqref{eq:VAinAdS}, i.e. $\CV_A^2 = \frac{1}{3}(\bar \CM-1)\bar m^2$, which was found through the combination of calculations in hydrodynamics (effective field theory from Section \ref{sec:LineFieldTheory}) and holographic thermodynamics in Section \ref{subsec:thermo}. 

We note that if we set $m=0$ in \eqref{PolyFiniteRho} and solve the cubic equation, then the three solutions contain the two diffusive modes with dispersion relations from Eqs. \eqref{M2diffusion} and \eqref{Full2solutionsHydro}, i.e. momentum and charge diffusion associated with the one-form symmetry, respectively. The third mode is the gapped, non-hydrodynamic mode \eqref{Full2solutionsNonHydro}. In other words, while the spectrum of the $\rho > 0 $ theory is consistent with that at $\rho = 0$, for the (two) hydrodynamic dispersion relations, the operations of taking the limit $\bar m\to 0$ and expanding around $ \bar k \approx 0$ do not commute (see also \cite{Grozdanov:2017kyl,Hernandez:2017mch}).

As a further check on the effective field theory set up in this work and on its consistency with the holographic dual studied here, we also compute the retarded transverse stress-energy two-point function from the bulk, finding
\begin{align}
G^{ty,ty}_{TT,R}(\omega,k) = \frac{\left(\bar \CM-1\right)\bar m^2  k^2}{\omega^2-\frac{1}{3}\left(\bar \CM-1\right)\bar m^2 k^2} + \ldots \,.
\end{align}
Comparing this result with the hydrodynamic Green's function from Eq. \eqref{eq:TtyTtyCorrelator} and using the thermodynamic quantities from Section \ref{subsec:thermo}, we see that it is not only pole (or the denominator of the Green's function) that is consistently computed by the two approaches but also the residue (or the numerator of $G^{ty,ty}_{TT,R}$).  

To uncover the regime of validity and limitations of the effective hydrodynamic theory from Section \ref{sec:Introduction}, we first note that as the dimensionless parameter $\bar m$ is tuned to become larger, $\bar m \in [0, \sqrt{6}]$, the temperature of the dual field theory state is tuned from $T = 3 r_h / 4 \pi$ to $T = 0$---the endpoint at which the bulk geometry becomes that of $AdS_2 \times \mathbb{R}_2$ with non-zero entropy density at zero temperature. A complete dual field theoretic construction of a low-energy effective theory of transport (hydrodynamics) near, and at the extremal point is a well-known and notorious problem due to the presence of a (large or infinite) set of low-energy modes on the imaginary $\omega$ axis. Despite this fact, certain features of the spectrum behave according to simple hydrodynamic predictions (see \cite{Edalati:2010pn,Davison:2013bxa}). In our case, a numerically computed speed of sound shows that the simple hydrodynamic prediction for the leading $\bar k$-dependent behaviour of the ``hydrodynamic" poles breaks down for large, non-perturbative $\bar m$, although this statement is very sensitive to the size of $\bar\CM$.

\begin{figure*}[t]
\begin{center}
\includegraphics[width=0.45\textwidth]{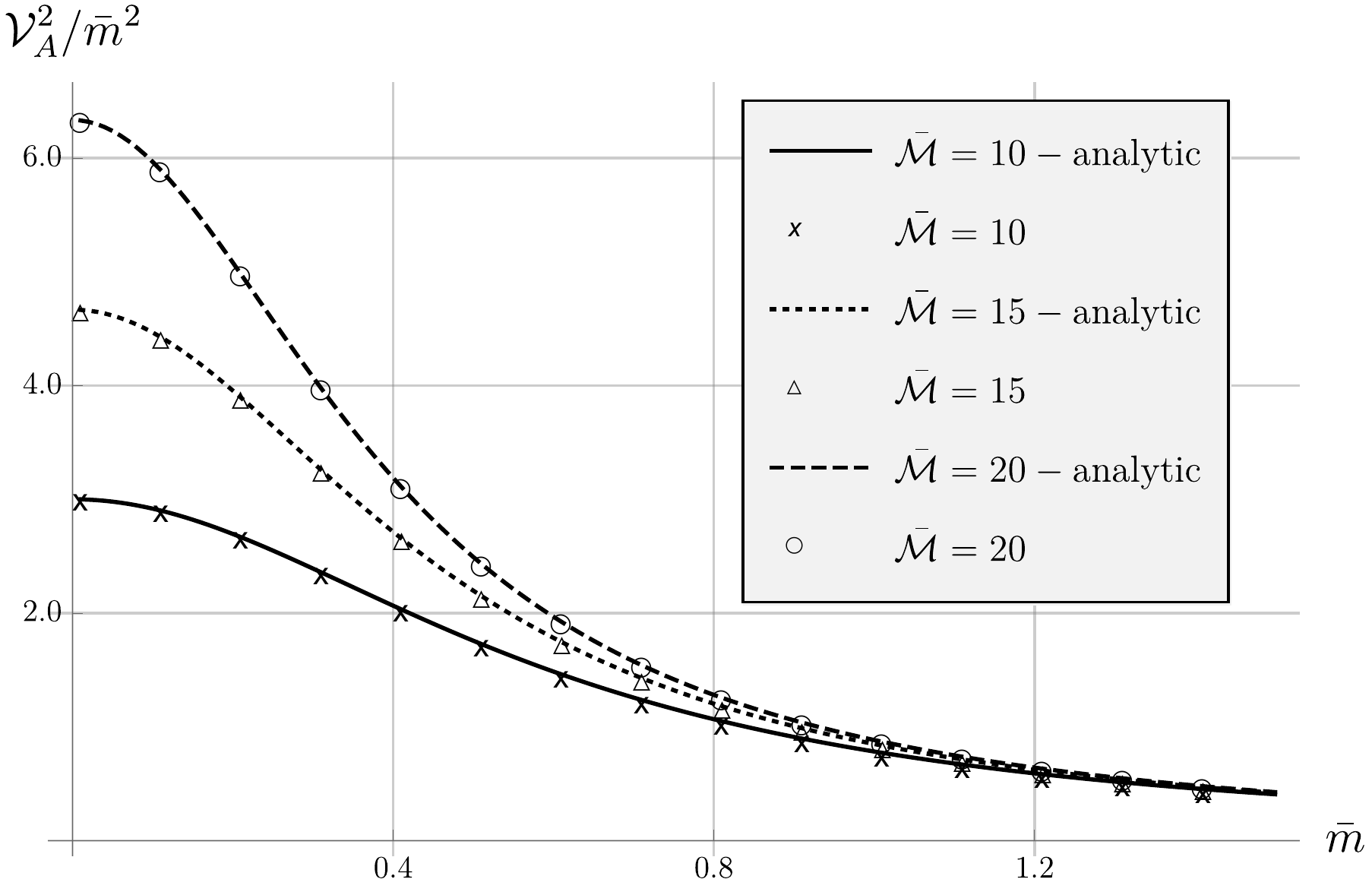} 
\hspace{0.05\textwidth}
\includegraphics[width=0.45\textwidth]{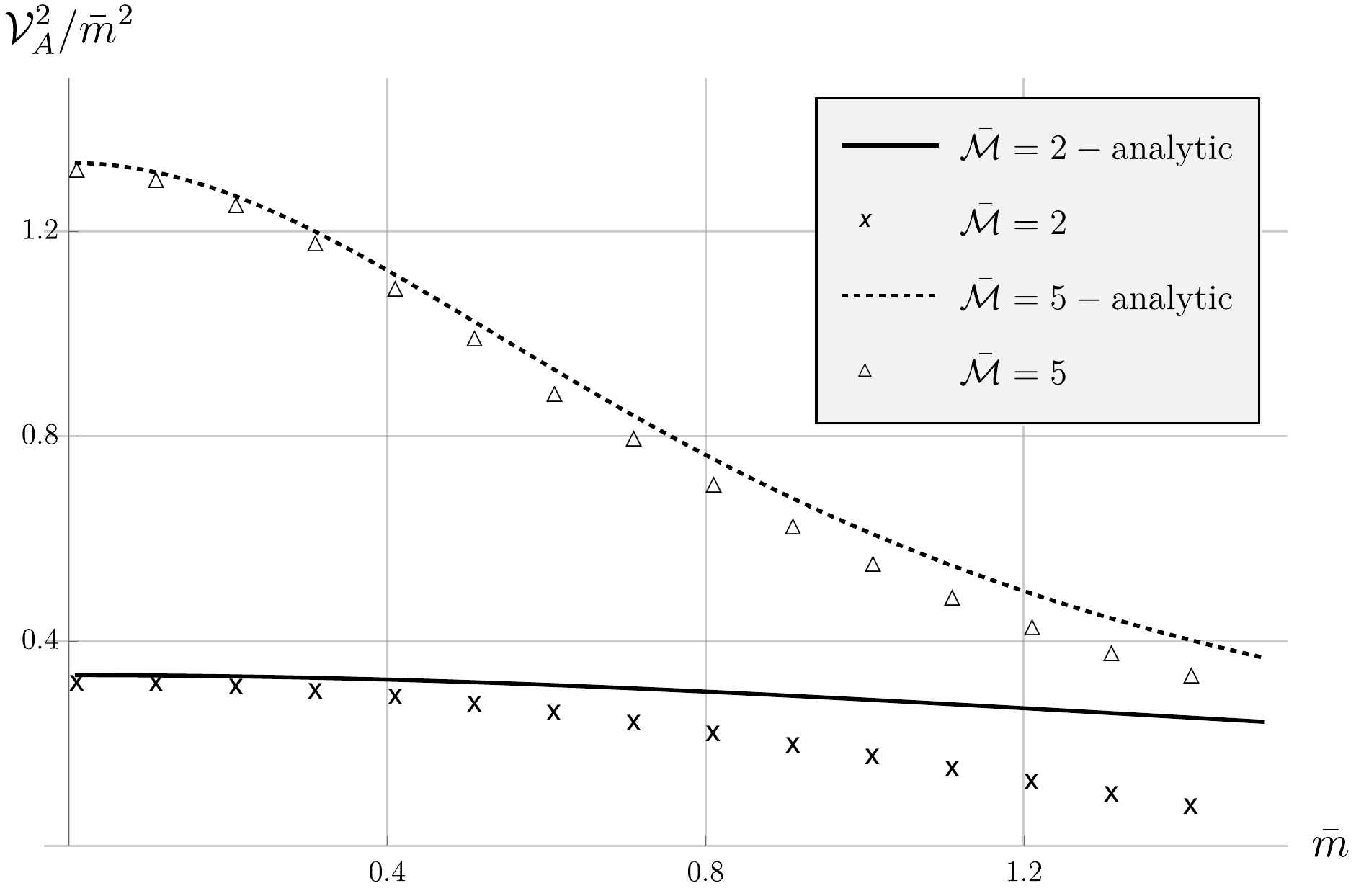}
\end{center}
\caption{Comparison between numerically computed speeds of transverse sound waves for $\bar\CM = \{2,5,10,15,20\}$ and the hydrodynamic, analytical result of Eq. \eqref{eq:VAinAdS}. The hydrodynamic prediction deviates from numerics for large values of $\bar m$. As is apparent from the plots, the agreement is better for larger values of the dimensionless parameter $\bar\CM$. }
\label{fig:sound-num}
\end{figure*}

Our numerical procedure again uses a small $\delta$-dependent expansion of $Z_\pm$ (cf. \eqref{eq:solHydroLim2}), with $\bar m$ kept arbitrary, but now expanded only to leading order in $\delta$, i.e. to $z_{0,\pm} (u,\omega,k,r_h,m) \delta^0$. We solve for $z_{0,\pm}$ numerically with the boundary conditions set by Eqs. \eqref{Z+PrescU}--\eqref{Z-PrescU}. While the agreement between numerical results and the hydrodynamic prediction \eqref{eq:VAinAdS} indeed becomes worse for larger $\bar m$ (as the extremal $T=0$ point is approached), it is interesting that increasing the renormalisation group scale $\bar \CM$, which enters the dispersion relations through the quasinormal mode prescription, significantly improves the hydrodynamic prediction. In fact, what we observe is that as $\bar\CM$ is tuned to be large---i.e. in the $\CM \gg T$ and $\CM \gg \rho$ regime, which is far from the instability that occurs below $\bar\CM = 1$---the agreement between the effective hydrodynamic prediction \eqref{eq:VAinAdS} and the speed of sound computed numerically becomes very good for large ranges of $\bar m$. The results are presented in Fig. \ref{fig:sound-num}. We leave a more detailed analysis behind the reasons for the breakdown of effective theory (detailed quasinormal mode spectrum, numerics done beyond the hydrodynamic expansion, etc.) to future works. Unlike in the higher-dimensional case of a hydrodynamic theory with a one-form symmetry (in MHD) where it was argued in Ref. \cite{Grozdanov:2016tdf} that there could exist a hydrodynamic effective theory at $T=0$ with the UV cut-off set by the density of the flux lines (the magnetic field), in this case, it is likely that this may not be possible. From the point of view of holography, it is plausible that such differences could be ascribed to the drastically different behaviour of duals to $AdS_2$ and $AdS_3$ geometries.

We end by noting that it would also be interesting to better understand the role $\bar\CM$ plays in potential instabilities in the spectrum when $\CM$ is small compared to other scales in the problem, e.g. when $\bar \CM < 1$ for small $\bar m$ (see discussions below Eqs. \eqref{ChemPot}, \eqref{eq:VAinAdS} and \eqref{eq:LatticeDiffu}).

\section{Conclusion}

In this work, we presented a new language for constructing effective field theories of states with fluctuating defects of arbitrary dimensionality. The central concept that enables such constructions are generalised global symmetries \cite{Gaiotto:2014kfa}, which were recently used to construct and extend the theory of long-range dynamics in plasmas, i.e. magnetohydrodynamics \cite{Grozdanov:2016tdf}. In MHD, the microscopic origin of a higher-form, generalised global symmetry is clear. In this paper, we argued that one can take the usefulness of such symmetries further and apply them to setups without a known microscopic, particle-oriented motivation, and apply them based purely on the basis of their topological and geometric content. The example we chose to study was a viscoelastic fluid with transverse sound modes---a subject which has recently attracted a considerable amount of attention in Refs. \cite{Amoretti:2016bxs,Esposito:2017qpj,Alberte:2017cch,Alberte:2017oqx,Amoretti:2017frz}, which include related constructions to the one presented in this work. Furthermore, in the second part of this paper, we showed how to construct holographic duals of such states (extending the works in Refs. \cite{Grozdanov:2017kyl,Hofman:2017vwr}) and performed a number of checks to show the consistency of the effective field theory with its dual and explored parameter regimes in which the boundary effective field theory breaks down for our particular setup. As is usual in holography, these duals should be thought of as UV completions of states, which flow to hydrodynamic effective theories in the IR. For this reason, holography enables us to study not only hydrodynamics, but also the interplay between hydrodynamic and non-hydrodynamic modes, as for example in Refs. \cite{Grozdanov:2016fkt,Grozdanov:2016vgg}. 

In future applications of this work, it would be interesting to systematically extend our construction to more geometrically complicated line defect structures (e.g. triangular or a hexagonal ``lattice" structure), include additional global symmetries or study scenarios in which the symmetries are discrete, anomalous or broken in a variety of phenomenologically-relevant ways. Furthermore, it would be interesting to systematically classify all hydrodynamic dissipative corrections to the constitutive relations from Section \ref{sec:Introduction}. Lastly, we believe that a wide variety of applications of generalised global symmetries may be found in condensed matter-motivated scenarios, related for example to Refs. \cite{Kivelson1998,PhysRevLett.84.753,PhysRevB.64.115105,Delacretaz:2016ivq,Delacretaz:2017zxd}.

\acknowledgements{We would like to thank Matteo Baggioli, Sasha Krikun, Jaakko Nissinen, Koenraad Schalm, Miggy Sulangi and Jan Zaanen for illuminating discussions. We are also very grateful to Blaise Gouteraux and Nabil Iqbal for their comments on a draft of this paper. S. G. is supported by the U.S. Department of Energy under grant Contract Number DE-SC0011090. N.P. is supported by DPST scholarship from the Thai government, Leiden University and Icelandic Research Fund grant 163422-052.}
    
\begin{appendix}
    
\section{The equilibrium partition function}\label{app:EQPF}

In this section, we derive the ideal constitutive relations and the thermodynamic relations stated in Eqs. \eqref{eq:DefIdealStessEn}--\eqref{Thermo2} by using the equilibrium partition function of \cite{Jensen:2012jh,Banerjee:2012iz}. Let us consider the state described in Section \ref{sec:Introduction} to be defined on a three-dimensional compact manifold with a (Euclidean signature) background metric $g_{\mu\nu}$ that has the topology of a three-torus. In equilibrium, the system possesses three independent Killing vectors $K^\mu_0$, $K^\mu_1$ and $K^\mu_2$. In the Wick-rotated Lorentzian space, $K^\mu_1$ and $K^\mu_2$ are spacelike and $K^\mu_0 \sim u^\mu$ is a timelike Killing vector that parametrises time translations. On $g_{\mu\nu}$, $K^\mu_0$ thus parametrises translations along the thermal cycle with the length of $\bar L_0$. Spacelike Killing vectors $K^\mu_{I}\sim h^\mu_{I}$ are directed along the line defects on cycles with lengths $\bar L_{I}$. Recall that $I \in \{1,2\}$. Furthermore, we take $K^\mu_0$ to be orthogonal to $K^{\mu}_{I}$ i.e. $g_{\mu\nu} K_0^\mu  K_{I}^\nu = 0$, which leads to the desired orthogonality conditions used in Section \ref{sec:Introduction}, i.e. $u_\mu h^\mu_1 = u_\mu h^\mu_2 = 0$. 
 
To make contact with the setup of this paper, we need to further couple the theory to two independent background two-form gauge fields $b_{I,\mu\nu}$ that can source the two-form conserved currents $J^{\mu\nu}_I$. Following the procedure of \cite{Jensen:2012jh,Banerjee:2012iz,Grozdanov:2016tdf}, the generating functional $Z[g_{\mu\nu},b_{I,\mu\nu}]$ is to be constructed from all diffeomorphism- and gauge-invariant scalars in the theory. At zeroth order in derivative expansion, these quantities are the proper lengths of the three cycles:
\begin{align}
L_Q = \frac{{\bar{L}}_Q}{\sqrt{g_{\mu\nu} K_{Q}^{\mu} K^\nu_Q }} ,\, 
\end{align} 
where $Q \in \{0, I\}$, and two Wilson surfaces $\CW_I$ that are constructed from $b_{I,\mu\nu}$: 
\begin{align}
\CW_I = \text{exp}\left[ 2\beta_0 L_I  K_0^{[\mu} K_I^{\nu]}  b_{I,\mu\nu}\right] \,,
\end{align}
where $\beta_0 = \bar{L}_0$ is the equilibrium temperature determined by the size of the imaginary time cycle. For the state of interest, we assume that the two Wilson surfaces are independent. Finally, the chemical potentials $\mu_I$ can be defined as
\begin{align}
\mu_I = \frac { \ln(\CW_I)}{L_0 L_I} \,. 
\end{align}
With these ingredients in hand, we can then follow Ref. \cite{Grozdanov:2016tdf} and recover Eqs. \eqref{eq:DefIdealStessEn}--\eqref{Thermo2}. 


\section{Longitudinal fluctuations}\label{appendix:Longitudinal}

\begin{figure}[tbh]
\begin{center}
\includegraphics[scale=0.6]{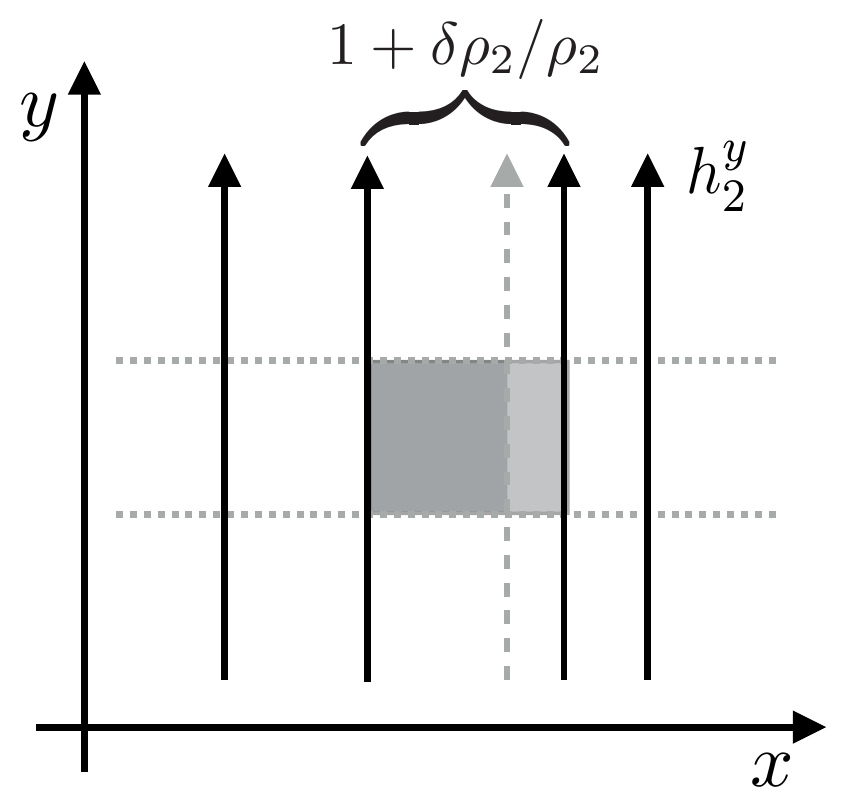}
\end{center}
\caption{The deformation of the lattice of lines by the transverse perturbation $\delta h_1^y(t,x)$. In a conventional elastic theory language, this corresponds to a distortion of a square (shaded area) due to a compressional deformation, i.e. the strain $U_{xx}$.}
\label{fig:Deformstrings-Long}
\end{figure}

In this part of the appendix, we complete the analysis of Sections \ref{sec:LineFieldTheory} and \ref{sec:HydroVsElasticity}, and show that the generalised global symmetry based formalism studied in this work agrees to linear order with the standard formulation of a fluid with elastic properties not only in the transverse, but also in the sound channel. Such fluctuations are even under the parity transformation $y \to - y$ with momentum pointing in the $x$-direction. In order to compare our computation to a conventional viscoelastic theory, it is convenient to consider the fluctuations of $\delta \rho_I$ as opposed to $\delta \mu_I$---i.e. we treat $\rho_I$ as hydrodynamic variables. First, the temporal components of the $\nabla_\mu J^{\mu \nu}_I = 0$ equations---the constraint equations $\nabla_\mu J^{\mu t}_I = 0$---imply that the following fluctuations must vanish:
\begin{align}
\delta \rho_1 = 0 \, , &&  \delta h^x_2 = 0 \,.
\end{align}
The only remaining non-trivial equation that follows from the conservation equations of the two-form currents is 
\begin{equation}
\partial_t \delta \rho_2 + \rho \, \partial_x \delta u^x = 0\,.
\end{equation}
In analogy with the transverse channel, this equation replaces the usual Josephson relation upon identification of $\delta U_{xx} = -\delta \rho_2/ \rho$, where $U_{xx} = \partial_x \phi_x$ is the strain tensor (cf. the definition below Eq. \eqref{eq:stressElastic}). Indeed, $\delta\rho_2/\rho$ precisely describes the compression of lines that are oriented perpendicularly to the direction of the propagation of momentum---i.e. the strain $U_{xx}$ action on a unit lattice cell (see Fig. \ref{fig:Deformstrings-Long}).

Next, our goal is to match the speed of the longitudinal sound of a particular defect configuration studied here to that derived from conventional hydrodynamics coupled to a scalar fields \cite{PhysRevA.6.2401,Delacretaz:2017zxd}. For this reason, we restrict our attention to the case of isotropic crystal structures. We start by computing the speed form the formalism based on generalised global symmetries. In the derivation, it is in fact convenient to relax the condition of isotropy, $\mu_1=\mu_2= \mu$ and $\rho_1 = \rho_2 =\rho$, and only impose it at the end. The conservation equations are now  
\begin{align}
\partial_t \delta \varepsilon + \chi_{p_xp_x} \partial_x \delta u^x & = 0 \,,\label{eq:conE}\\
\partial_t \delta p^x + \alpha_1 \partial_x \delta s - \alpha_2 (\partial_x \delta\rho_2)/\rho_2 & = 0 \, ,\label{conP}
\end{align}
where $\delta \varepsilon  =  T\delta s + \mu_1 \delta\rho_1 +  \mu_2 \delta\rho_2$, $\chi_{p_xp_x} = \varepsilon+p-\mu_1\rho_1$ is the momentum susceptibility and $\delta p^x = \chi_{p_xp_x} \delta u^x$ and spatial momentum. The functions $\alpha_1$ and $\alpha_2$ are combinations of thermodynamics variables and susceptibilities, namely, 
\begin{align}
\alpha_1 &= \left( \frac{\partial (p-\mu_1\rho_1)}{\partial s} \right)_{\rho_2},\\
\alpha_2 &=    -\rho_2\left(\frac{\partial p}{\partial\rho_2} \right)_s + \rho_1\rho_2 \left( \frac{\partial \mu_1}{\partial\rho_2} \right)_s \,. 
\end{align}

These expressions can now be compared with the equations of motion derived from the standard description of a viscoelastic medium from Section \ref{sec:HydroVsElasticity}, which uses the stress-energy tensor stated in Eq. \eqref{eq:stressElastic}. The conservation of energy immediately reduces to Eq. \eqref{eq:conE}. On the other hand, the conservation of momentum, $\partial_t \delta p^x + \partial_i \delta T^{ix} = 0$, gives
\begin{align}\label{eq:ElasticconP}
\partial_t \delta p^x  + \beta_1 \partial_x\delta s + \beta_2 \partial_x \delta U_{xx} = 0 \,,
\end{align}
where the coefficients $\beta_1$ and $\beta_2$ are
\begin{align}
\beta_1  &= \left( \frac{\partial(p_0-2\CB)}{\partial s}\right)_{U_{xx}} \,,\\
\beta_2 &= \left( \frac{\partial (p_0-2\CB)}{\partial U_{xx}} \right)_s - (\CB+\CG) \, .
\end{align}
Using the identification between the two descriptions from the transverse channel, i.e. Eq. \eqref{eq:Map2FormToElastic}, one immediately finds that $\alpha_1 = \beta_1$. As for $\beta_2$, we see that by using Eq. \eqref{eq:Map2FormToElastic} and $U_{xx} = -\delta\rho_2/\rho$, which follows from the Josephson relation in the longitudinal channel, $\beta_2$ becomes 
\begin{align}
\beta_2 = -\rho_2 \left(\frac{\partial p}{\partial \rho_2}\right)_s + \rho_2 \left(\frac{\partial \mu_2\rho_2}{\partial\rho_2} \right)_2- \mu_1\rho_1 - \CB \, .
\end{align}
Finally, we can deduce that $\alpha_2=\beta_2$ if the bulk modulus is identified with the following expression:
\begin{align}
\CB = \mu_2\rho_2 - \mu_1\rho_1+ \rho_2^2 \left( \frac{\partial\mu_2}{\partial \rho_2} \right)_s- \rho_1\rho_2 \left( \frac{\partial\mu_1}{\partial \rho_2} \right)_s \,.
\end{align}
Hence, under such identifications, the two descriptions of linear fluctuations match in both transverse and longitudinal channels. Note that in an isotropic state with $\mu_1=\mu_2= \mu$, $\rho_1 = \rho_2 =\rho$ and $(\partial \mu_2/\partial\rho_2)_s =(\partial\mu_1/\partial\rho_2)_s$, the bulk modulus $\CB$ vanishes.

\section{Renormalisation group conditions and the Wick rotation}\label{appendix:AnalyticCont}
In this part of the appendix, we discuss in more detail the analytic continuation of correlators from Euclidean to Lorentzian frequency and their associated renormalisation group condition, which was used in Section \ref{sec:zerowallzerotem}. In holography, the necessary procedure for performing the Wick rotation can be done by using the real-time prescription for holographic renormalisation of Refs. \cite{Skenderis:2008dh,Skenderis:2008dg}. 

Inspired by the Schwinger-Keldysh (closed-time-path) formalism \cite{schw,Bakshi:1962dv,Mahanthappa:1962ex,Keldysh:1964ud} (see also \cite{kamenev,calzetta,Grozdanov:2015nea}), let us consider a bulk theory on a complex time contour $C$, which connects a spacetime manifold with real time, $M_L$, and another with Euclidean time, $M_E$. The holographic generating functional of our bulk theory with $G_{ab}$ and $B_{I,ab}$ then becomes
\begin{equation}
\ln Z_C[g_{\mu\nu},b_{\mu\nu}]  = iS_L - S_E\, .
\end{equation}
The Lorentzian part of the on-shell action $S_L$, keeping track of only the higher-form bulk fields, can be written as
\begin{align}
S_L \supset& -\frac{1}{12}\int_{M_L} dt  d^3x \sqrt{-G}\, H_{abc}H^{abc}  \nn
&+\frac{1}{4\kappa_L(\Lambda)} \int_{\partial M_L} dt d^2x \sqrt{-\gamma}\, \CH_{\mu\nu} \CH^{\mu\nu}\Big\vert_{r = \Lambda}\, ,
\end{align}
and, similarly, the Euclidean part $S_E$ as 
\begin{align}
S_E \supset&\,\, \frac{1}{12}\int_{M_E} d\tau  d^3x \sqrt{G}\, H_{abc}H^{abc}  \nn
&\,\,-\frac{1}{4\kappa_E(\Lambda)} \int_{\partial M_E} d\tau d^2x \sqrt{\gamma}\, \CH_{\mu\nu} \CH^{\mu\nu} \Big\vert_{r = \Lambda}\, ,
\end{align}
where $\kappa_E(\Lambda)$ and $\kappa_L(\Lambda)$ satisfy the renormalisation conditions discussed in Section \ref{sec:HolRG}. Consequently, one can write the sources in the boundary field theory (cf. Eq. \eqref{eq:defSource}) on the Lorentzian and the Euclidean time segments of the contour $C$ as 
\begin{equation}
\begin{aligned}
b^L_{I,\mu\nu} &= \frac{1}{2} \left(\Lambda \, \CJ^L_{I,\mu\nu} +  \hat B^L_{I,\mu\nu} \right) - \frac{\Lambda}{2 \kappa_L}  \CJ^L_{I,\mu\nu}   \\
b^E_{I,\mu\nu} &= \frac{1}{2}  \left(\Lambda \, \CJ^E_{I,\mu\nu} +  \hat B^E_{I,\mu\nu} \right) - \frac{\Lambda}{2 \kappa_E}  \CJ^E_{I,\mu\nu}  ,
\end{aligned}
\end{equation}
with the running coupling on the two contours given by
\begin{align}
\frac{\Lambda}{\kappa_L} = \Lambda-\CM_L\,,&&  \frac{\Lambda}{\kappa_E} = \Lambda-\CM_E\, .
\end{align}
I.e., $\CM_L$ and $\CM_E$ can be two independent integration constants.

Up until this point, $M_L$ and $M_E$, and associated $\CM_L$ and $\CM_E$, were considered as independent manifolds and energy scales. However, in order for us to be able to analytically continue from one to another, we consider the behaviour of the theory around the point where $M_L$ and $M_E$ are glued together. Following Section 3.1 of \cite{Skenderis:2008dg}, one is required to impose a matching condition on the fields $B^L_{I,ab}$ and $B^E_{I,ab}$ across the point where the bulk manifolds $M_L$ and $M_E$ are glued together. Namely, we set
\begin{equation}\label{eq:matchingLEbulk}
B^L_{I,ab} = B^E_{I,ab} 
\end{equation}
at the boundaries of the two manifolds. Furthermore, for the variation of the bulk part of $\ln Z_C$ to be well-defined at this point---i.e. so that $\delta (iS_L-S_E) = 0$---the bulk field strengths must satisfy 
\begin{equation}\label{eq:matchingLEmom}
i H_{I,tab}^L + H_{I,\tau ab}^E = 0 
\end{equation}
at the point where $M_L$ and $M_E$ are glued together. Note that the matching condition \eqref{eq:matchingLEbulk} also implies that $\hat B^E_{I,ab} = \hat B^L_{I,ab}$ at the same point. Finally, it is clear from the Schwinger-Keldysh formalism that these conditions parallel the usual boundary conditions by which one is required to match fields at the ends of various parts of the time contour. In our case, the field theory sources should satisfy $b^L_{I,\mu\nu} = b^E_{I,\mu\nu}$ at the glueing point. Together, this condition and the matching conditions \eqref{eq:matchingLEbulk} and \eqref{eq:matchingLEmom} imply that one has to set $\CM_E =  i \CM_L$. Hence, the analytic continuation of the correlator $\la J^{\tau i}_I J^{\tau j}_I\ra$ to $\la J^{t i}_I J^{t j}_I\ra_R$ in Section \ref{sec:zerowallzerotem} requires us to use $\CM_E = i \CM$.
\end{appendix}

\bibliographystyle{utphys}
\bibliography{biblio}
\end{document}